%% file: rqe_main.tex
\documentclass{article}

\usepackage{arxiv}

\usepackage[utf8]{inputenc} 
\usepackage[T1]{fontenc}    
\usepackage{booktabs}       
\usepackage{amsfonts}       
\usepackage{nicefrac}       
\usepackage{microtype}      
\usepackage{lipsum}
\usepackage{cite}
\usepackage{amsmath,amssymb,amsfonts}
\usepackage{algorithmic}
\usepackage{graphicx}
\usepackage{textcomp}
\usepackage{multirow}
\usepackage{multicol}
\usepackage{xcolor}

\def\BibTeX{{\rm B\kern-.05em{\sc i\kern-.025em b}\kern-.08em
    T\kern-.1667em\lower.7ex\hbox{E}\kern-.125emX}}
\begin{document}

\title{DESIGN OF SECURE AND ROBUST COGNITIVE SYSTEM FOR MALWARE DETECTION\\
}

\author{Sanket Shukla\\
Electrical and Computer Engineering \\
\textit{George Mason University}\\
Fairfax, USA. \\
sshukla4@gmu.edu}

\maketitle

\begin{abstract}
    The computer systems for decades have been threatened by various types of hardware and software attacks of which Malware have been one of the pivotal issues. This malware has the ability to steal, destroy, contaminate, gain unintended access, or even disrupt the entire system. There have been techniques to detect malware by performing static and dynamic analysis of malware files, but, stealthy malware has circumvented the static analysis method and for dynamic analysis, there have been previous works that propose different methods to detect malware. However, these techniques do not perform well on stealthy malware.  Moreover, the rising trend and advancements in machine learning has resulted into its numerous applications in the field of computer vision, pattern recognition to providing security to hardware devices. Machine learning based malware detection techniques rely on grayscale images of malware and tends to classify malware based on the distribution of textures in graycale images. Albeit the advancement and promising results shown by machine learning techniques, attackers can exploit the vulnerabilities by generating adversarial samples. Adversarial samples are generated by intelligently crafting and adding perturbations to the input samples. There exists majority of the software based adversarial attacks and defenses. To defend against the adversaries, the existing malware detection based on machine learning and grayscale images needs a preprocessing for the adversarial data. This can cause an additional overhead and can prolong the real-time malware detection. So, as an alternative to this, we explore RRAM (Resistive Random Access Memory) based defense against adversaries. Therefore, the aim of this thesis is to address the above mentioned critical system security issues. The above mentioned challenges are addressed by demonstrating proposed techniques to
design a secure and robust cognitive system. 
First, a novel technique to detect stealthy malware is proposed. The technique uses malware binary images and then extract different features from the same and then employ different ML-classifiers on the dataset thus obtained. Results demonstrate that this technique is successful in differentiating classes of malware based on the features extracted.
Secondly, I demonstrate the effects of adversarial attacks on a reconfigurable RRAM-neuromorphic architecture with different learning algorithms and device characteristics. I also propose an integrated solution for mitigating the effects of the adversarial attack using the reconfigurable RRAM architecture.
\end{abstract}

\input{chapterOne}

\input{chapterTwo}

\input{ChapterThree}

\input{conclusion}

\input{Future}
\bibliographystyle{IEEEtran}
\bibliography{malware_ait624}
\end{document}

%% file: chapterOne.tex
\section{Introduction to Malware Threats}
The hardware security discipline in recent years experienced a
plethora of threats like the Malware attacks \cite{Dhavlle_DATE'21, Dhavlle_ISCAS'21, Meraj_ISCAS'21, SMPD_DAC'19, Sanket_CASES'19, Sanket_ICMLA'19, Sanket_ICTAI'19}, Side-Channel Attacks \cite{Yarom_usenix_'14,Gruss_dimva_'16, Abhijitt_isqed'20, Dhavlle_TCAD'21}, Hardware Trojan attacks \cite{Meraj_AsianHOST'20}, reverse engineering threats \cite{Kolhe_GLSVLSI'19, Kolhe_ICCAD'19, Hassan_ISQED'20, dac2021, dac2022_2, dac2022, myjournal, frontiers, DATE, raki1, raki2, vivek1, vivek2019, vivek2020} and so on. I focus on the malware detection technique here along with some state-of-the-art works.
Malicious Software, generally known as `malware' is a software program developed by an attacker to gain unintended access of a computer system for performing illegit and malicious activities like stealing data, sensitive information  ( like passwords, SSN's), contaminating and manipulating data without users consent. According to 2018 threat report by McAfee labs, about 73 million malicious files, 66 million malicious IP addresses and and 63 million malicious URL's were detected and declared as risky. Similar threat report by McAfee reported 57.6 million malicious files in 2017. This rapidly increasing trend of generation of malware is a serious threat and global concern for the community. It results in need to develop a promising and comprehensive malware detection methods with robustness. Traditional and primitive software based solutions for malware detection methods such as signature-based and semantics-based anomaly detection techniques induce remarkable computational overheads to the system \cite{sign_based_technique_1} \cite{sign_based_technique_2} \cite{sign_based_technique_3} \cite{sign_based_technique_4}.
Traditional approaches towards analyzing malware involve extraction of binary signatures from malware, comprising their fingerprint. There is an exponential increase in the number of new signatures released every year, due to the rapid escalation of malware.

\subsection{Existing Machine Learning and Deep Learning Defense Against Malware }

Machine learning is an emerging technique and is extensively used in various fields like computer-vision, pattern-recognition, natural language processing, computer security etc. where the massive volume of data is generated regularly. Among several machine learning classifiers, the neural network class, especially deep neural networks (DNNs) and convolutional neural networks (CNNs), have tremendously transformed the capabilities and computational power of the computer systems. Some of the machine learning applications in the aforementioned fields includes self-driving cars \cite{self_driving_1}, deep space exploration, weather prediction, object recognition, and so on. Advancements and progress in the field of computer-vision has anticipated the development of self-driving cars without any human intervention \cite{self_driving_survey}. Similarly, machine learning has shown promising results to secure computer systems against malware and stealthy malware using image recognition \cite{shukla_wip} and pattern recognition \cite{Sanket_ICTAI'19, sanket_dac21} techniques. Despite the benefits and results showcased by advancements in machine learning \cite{sreenitha_1}, the existing vulnerabilities tend to exploit by impacting the performance of the machine learning classifier.

\subsection{Introduction to Adversarial Training Data}

Although the machine learning techniques tend to be robust to the noise, the exposed vulnerabilties has shown that the output of machine learning classifier can be easily manipulated by crafting perturbations to the input data \cite{szegedy2013intriguing, goodfellow2014explaining, papernot2016limitations}. The data generated by crafting perturbations is generally known as Adversarial samples. These adversarial samples are constructed by perturbing the input data in one or multiple
cycles iteratively under certain constraints in order to escalate the classification error rate.

\begin{figure}[!htb]
    \figSpace
    \centering
    \includegraphics[width=0.9\textwidth]{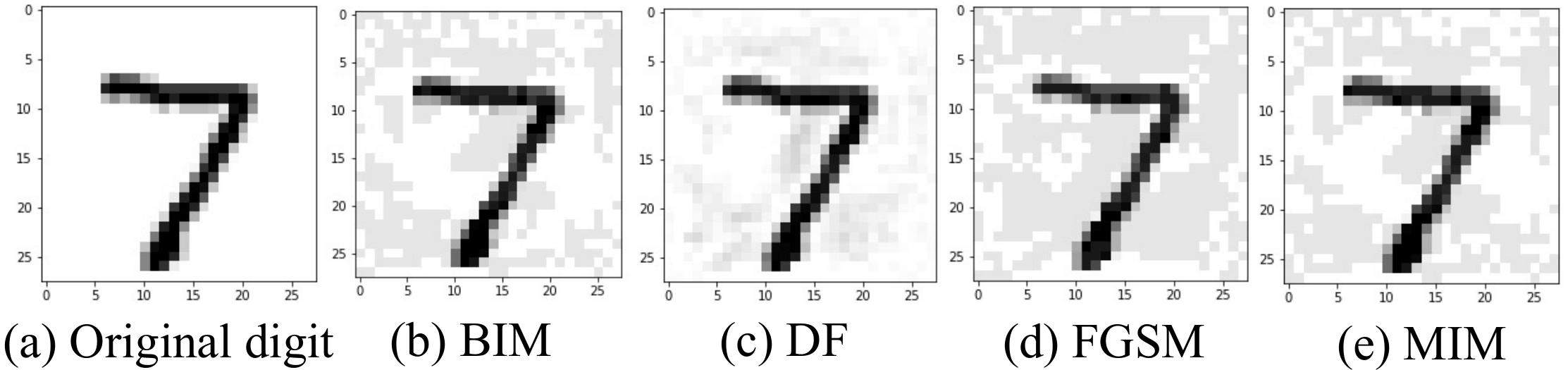} 
    \caption{ (a) Original MNIST digit ``7"; (b) BIM generated adversarial image; (c) DF generated adversarial image; (d) FGSM generated adversarial image and (e) MIM generated adversarial image } 
    \figSpace
    \label{fig:adv_im}
\end{figure}

Figure \ref{fig:adv_im} illustrates a simple adversarial sample generated from the MNIST digit dataset \cite{mnist} for digit ‘7’. The Figure \ref{fig:adv_im}(a) is the orgiginal image which is classified as 7 by the neural network classifier. The images in Figure \ref{fig:adv_im}(b), \ref{fig:adv_im}(c), \ref{fig:adv_im}(d), \ref{fig:adv_im}(e) are generated by the  basic iterative method (BIM), Deepfool attack (DF), fast gradient sign method (FGSM), and momentum iterative method (MIM) respectively. It can observed  from the Figure \ref{fig:adv_im}(a), \ref{fig:adv_im}(c) and \ref{fig:adv_im}(e) that the normal and adversarial samples look similar for human observation. It needs to be noted that the noise in Figure \ref{fig:adv_im}(c) and \ref{fig:adv_im}(e) can be increased or reduced by tuning the parameters of the attack. With the change in attack parameters, the classifier output and it’s confidence will be modified. More details on generating the adversarial attacks are presented in Section \ref{attacks and defense}.

In this work, the intent is to design a secure cognitive system by proposing two defense mechanisms to mitigate the malware and adversarial attacks. The proposed defenses are organized in two distinct chapters and discussed in detail.

%% file: chapterTwo.tex
\section{Malware Detection by Extracting Gray-scale Image Features}

As seen in the introduction, static code analysis and dynamic code analysis are some of the approaches used for analyzing malware. Static analysis looks for malicious patterns by disassembling the code and exploring the control flow of the executable. Whereas, in  dynamic analysis malicious code is executed in a virtual environment and based on the execution trace a behavioral report characterizing the executable is generated. These techniques have their pros and cons. Although, static analysis offers the most complete coverage but it suffers from code obfuscation. Prior to analysis, the executable has to be unpacked and decrypted, and even then, the analysis can be thwarted by problems of intractable complexity.
Dynamic analysis does not need the malware executable to be unpacked or decrypted and is also more efficient but it is time intensive and consumes resources, which results in scalability issues. Moreover, sometimes environment does not satisfy the triggering conditions, leaving some malicious behaviors unobserved.

This work focuses on a completely different and a novel approach to characterize and analyze malware. Approach tends to represent a malware executable as a binary string of zeros and ones. Furthermore, this vector of binaries can be reshaped and converted into a matrix which can be later viewed as an image. Malware belonging to same family showed significant visual similarities in image texture. We discuss representing malware binaries as images. We consider malware classification problem as one of image classification problem. Existing classification techniques require either disassembly or execution whereas our method does not require either but still shows significant improvement in terms of performance. Moreover, our method is also resilient to popular obfuscation techniques such as section encryption. This automatic classification technique should be very valuable for anti-virus companies and security researchers who report thousands of malware everyday.

\subsection{Literature Review and Motivation}

Several tools such as text editors and binary editors can both visualize and manipulate binary data. There have been several GUI-based tools which facilitate comparison of files. However, there has been limited research in visualizing malware. In \cite{5_Yoo_2004} Yoo used self organizing maps to detect and visualize malicious code inside an executable. In \cite{6_quist} Quist and Liebrock develop a visualization framework for reverse engineering. They identify functional areas and de-obfuscate through a node-link visualization where nodes represent the address and links represent state transitions between addresses. In \cite{7_trinius} Trinius et al. display the distributions of operations using treemaps and the sequence of operations using thread graphs. In \cite{8_Goodall} Goodall et al. develop a visual analysis environment that can aid software developers to understand the code better. They also show how vulnerabilities within software can be visualized in their environment.

While there hasn’t been much work on viewing malware as digital images, Conti et al. \cite{9_Conti} visualized raw binary data of primitive binary fragments such as text, C++ data structure, image data, audio data as images. In \cite{10_Conti} Conti et al. show that they can automatically classify the different binary fragments using statistical features. However, their analysis is only concerned with identifying primitive binary fragments and not malware. This work presents a similar approach by representing malware as grayscale images.

Several techniques have been proposed for clustering and classification of malware. These include both static analysis \cite{11_Karim05malwarephylogeny,12_Kolter,13_Gao,14_Tian,15_tian,16_Tian,17_Gheorghescu2006ANAV} as well as dynamic analysis \cite{18_park,19_Bailey:2007,20_Bayer_scalable,21_Rieck:2008}. We will review papers that specifically deal with classification of malware. In \cite{21_Rieck:2008} Rieck et al. used features based on behavioral analysis of malware to classify them according to their families. They used a labeled dataset of 10,072 malware samples labeled by an anti-virus software and divide the dataset into 14 malware families. Then they monitored the behavior of all the malware in a sandbox environment which generated a behavioral report. From the report, they generate a feature vector for every malware based on the frequency of some specific strings in the report. A Support Vector Machine is used for training and testing the feature on the 14 families and they report an average classification accuracy of 88\%.
In contrast to \cite{21_Rieck:2008}, Tian et al \cite{14_Tian} use a very simple feature, the length of a program, to classify 7 different types of Trojans and obtain an average accuracy of 88\%. However, their analysis was only done on 721 files. In \cite{15_tian,16_Tian} the same authors improve their above technique by using printable string information from the malware. They evaluated their method on 1521 malware consisting of 13 families and reported a classification accuracy of 98.8\%.
In \cite{18_park}, Park et al. classify malware based on detecting the maximal common sub graph in a behavioral graph. They demonstrate their results on a set of 300 malware in 6 families.

With respect to related works \cite{Abhijitt_1, Abhijitt_2, Abhijitt_3, Abhijitt_4,Abhijitt_5, Abhijitt_6, Abhijitt_7 }, our classification method does not require any disassembly or execution of the actual malware code. Moreover, the image textures used for classification provide more resilient features in terms of obfuscation techniques, and in particular for encryption. Finally, we evaluated our approach on a larger dataset consisting in 25 families within a malware corpus of 9,458 malware. The evaluation results show that our method offers similar precision at a lower computational cost.

\subsection{EDA Analysis of the Malign Dataset}

\begin{figure*}
    \figSpace
    \centering
    \includegraphics[width=1\textwidth]{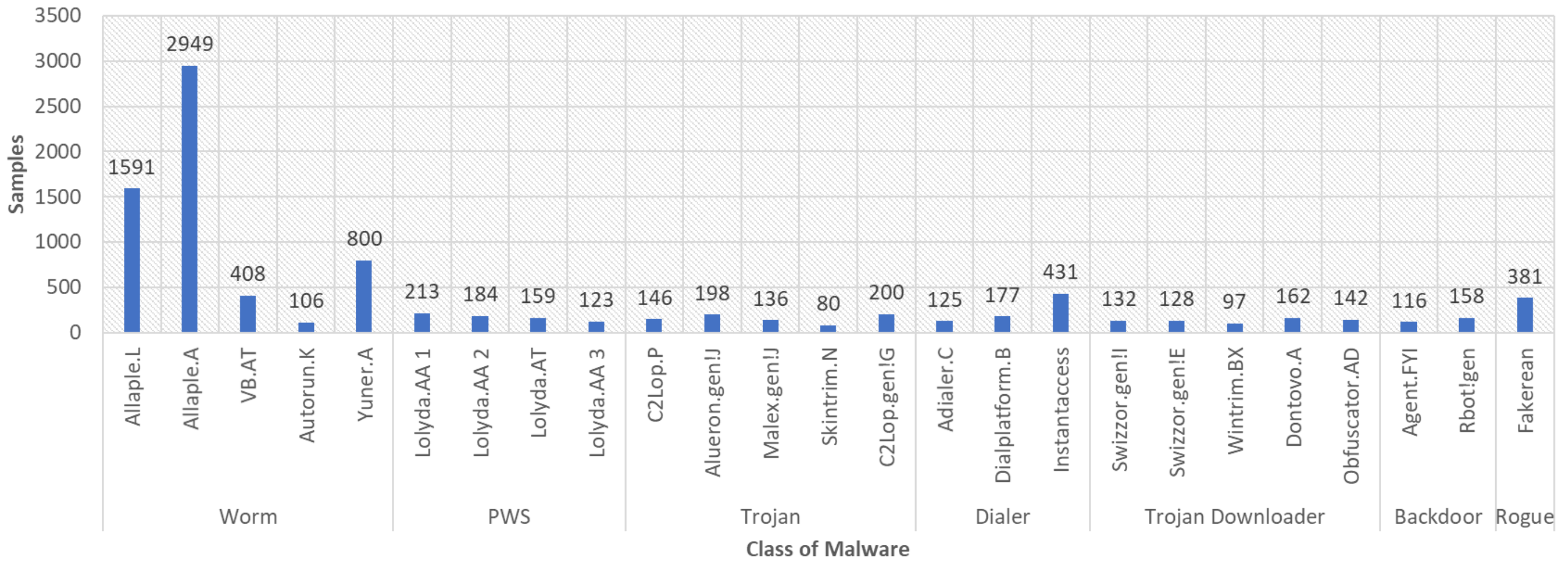}
    \vspace{10pt}
    \caption{Dataset distribution with Classes and Sub-Classes \vspace{10pt}}
    \figSpace
    \label{fig:malimg}
\end{figure*}

To mitigate the issue of classification of malware our first step was to perform exploratory data analysis (EDA) on the dataset. We are using malimg dataset for the analysis and the dataset distribution is as shown in Figure \ref{fig:malimg}. Prior to perform Exploratory data analysis we extracted some important features to create training and testing dataset. We extracted the following features for grayscale images in dataset: ``energy, entropy, contrast, dissimilarity, homogeneity and correlation". ``Entropy" defines statistical measure of randomness used to characterize the texture of the input image. ``Energy" defines sum of squared elements in the gray level co-occurence matrix. ``Contrast" defines intensity contrast between pixel and neighbor. ``Dissimilarity" degree of dissimilarity between images. ``Homogeneity" measures the closeness of the distribution of elements in the gray level co-occurence matrix to gray level co-occurence matrix  diagonal. ``Correlation" defines correlation between a pixel and its neighbor over entire image.

\begin{figure}[!htb]
    \figSpace
    \centering
    \includegraphics[width=0.45\textwidth]{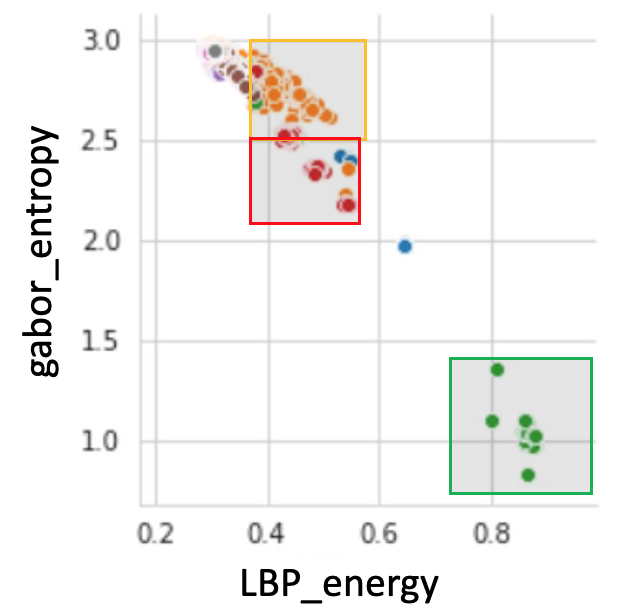}
    \caption{Gabor-entropy v/s LBP-energy}
    \label{fig:eda_1}
    \figSpace
\end{figure}

\begin{figure}[!htb]
    \figSpace
    \centering
    \includegraphics[width=0.45\textwidth]{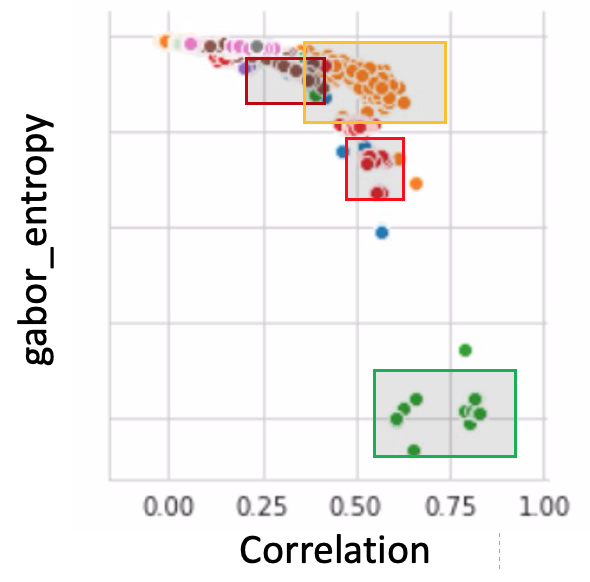}
    \caption{Gabor-entropy v/s Correlation}
    \label{fig:eda_2}
    \figSpace
\end{figure}

Figure \ref{fig:eda_1} shows 2-D scatter plot of gabor-entropy v/s LBP-energy. Here we can easily cluster datapoints of 3 different classes, however there are many overlapping datapoints which makes it difficult to rely on these two features for malware classification. Similar when we select other features as shown in Figure \ref{fig:eda_2} gabor-entropy v/s correlation. In this 2-D scatter plot we can cluster datapoints of 4 different classes but still some datapoints are overlapped. So, neither of the 2-D scatter plots could give us significant classification outcomes. The graphs in Figure \ref{fig:eda_2} and Figure \ref{fig:eda_1} exhibits collinearity and to overcome this we extracted new features by performing feature engineering on current features by applying some mathematical functions like log, square, cube, etc. To classify the datapoints in the overlapping reqion was a big challenge. At the same time we had to consider time complexity, power and performance factors as well. This calls for more robust malware detection mechanism.

\begin{itemize}
    \item Here explain all the extracted features in detail - like one small paragraph each feature and how you extracted them
    
    \item Mention how you converted the binaries to the images.
    
    \item Any challenges you faced. Then how you extracted the features.
\end{itemize}

\subsection{Experimental Setup and Data Collection}

All the data collection was done using Python scripts for different classes of malware images and the extracted features. The experiments were run on a Windows 10 OS with Intel i7 Coffeelake, 32GB RAM, NVidia GTX 1080 Ti 8GB Graphics Card. The scripts were run in Jupyter notebook and the machine learning models were trained on the same machine with Weka tool \cite{Hall_weka'09}.

\subsection{Proposed Technique for Malware Detection}

\begin{figure*}[!htb]
    \figSpace
    \centering
    \includegraphics[width=1\textwidth]{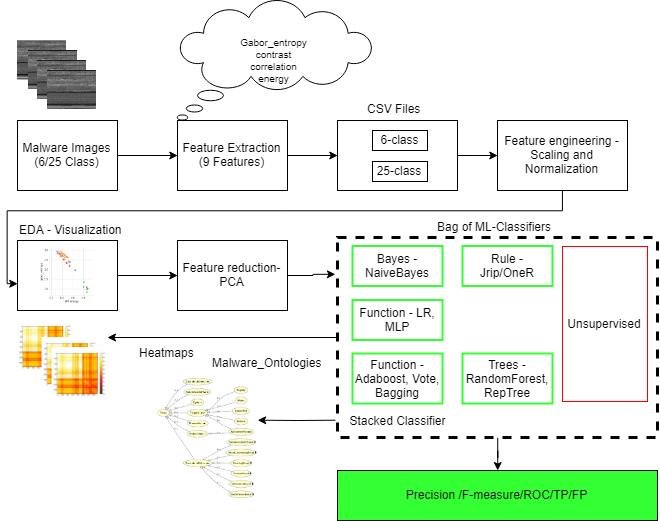}

    \caption{Dataset distribution with Classes and Sub-Classes}
    \label{fig:process_diagram}
    \figSpace
\end{figure*}

Figure \ref{fig:process_diagram} depicts the entire process that we have performed to detect malwares and categorize them in different classes.The first three blocks of the Figure \ref{fig:process_diagram} namely Malware Images, Feature Extraction, constructing CSVs and EDA visualization have been discussed in detail. Here we describe the other blocks of the diagram. 
So far we have learned all the background required for detecting malware using features corresponding to the malware images which were extracted from malware binaries. The conclusion of our EDA motivated that further analysis needed to be done to make fruitful use of the dataset created and process it in such a way that would give better and desired results after classification. The whole purpose is to prepare the data such that we do not need to tweak the classification models to a great extent and hence save a lot of time. 
Until now we are done with:

\begin{itemize}
    \item Obtaining the dataset
    \item Converting binaries to grayscale images
    \item Developing Python scripts to extract different features related to images from the dataset
    \item Generating CSV dataset files for the 6-main classes and 25-sub classes of malware
    \item Exploratory data analysis for dataset visualization of the 6-main and 25-sub classes to conclude what direction to proceed in
\end{itemize}

Depending upon the EDA results, we were suggested to go for a feature selection technique that basically reduces our features to avoid irrelevant features being taken into consideration. 
\subsubsection{Feature Reduction and Selection using Principle Component Analysis} There are many feature reduction and selection algorithms available but we decided to go with the most common PCA (Principle Component Analysis) algorithm. As the name suggests, PCA tries to minimize dataset to relevant features. There are many reasons why PCA component analysis is crucial before training classifiers:

\begin{itemize}
    \item Reducing Computation Time and Complexity: More the features, more time required to train the ML-model. Hence less complexity of the implemented model.
    \item Remove Irrelevant Features: There might be many features in the dataset that do not significantly contribute to 'relevant information' needed for best classification accuracy results, hence, removing or reducing the dataset to the most relevant ones will provide better accuracy under less time.
\end{itemize}

PCA is an algorithm that constructs totally new features known as Principle Components (PCs). These newly constructed features basically cover all the variance and information of the dataset thus reducing the complexity of models without compromising the information in the dataset. We had set the configuration in the Weka tool to include 95\% of the information of the dataset (both 6-class and 25-subclasses) with at the most 5 features while building the PC equations. The results of our PCA analysis is discussed in the Section \ref{results}. 

\subsubsection{{Feature Engineering - Scaling and Normalization}} 
\begin{figure*}
    \figSpace
    \centering
    \includegraphics[width=0.9\textwidth]{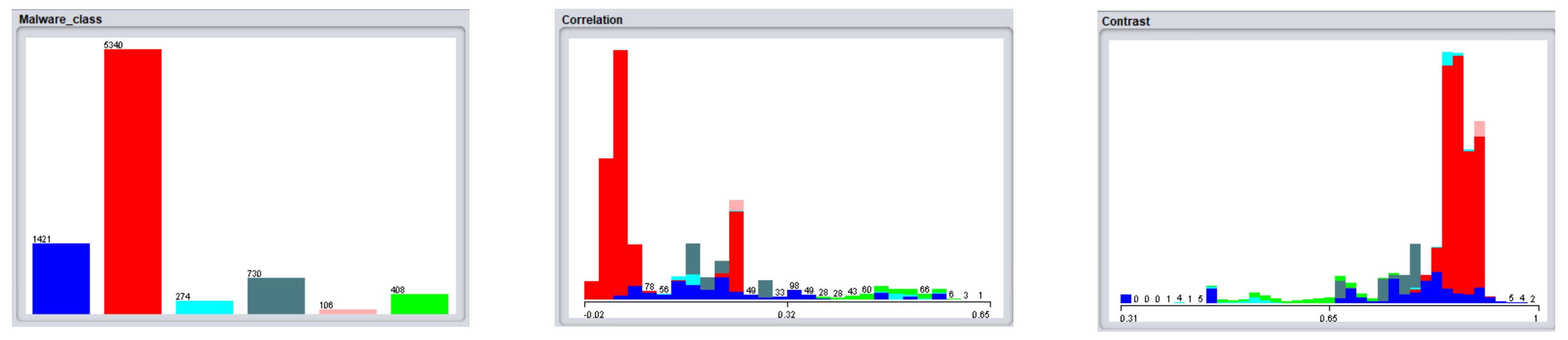}
    \vspace{1em}
    \caption{Visualization for 6-main classes; (a) All malware classes, (b) Correlation feature distribution and (c) Contrast feature distribution}
    \label{fig:visu_six}
    \figSpace
\end{figure*}

\begin{figure*}
    \figSpace
    \centering
    \includegraphics[width=0.9\textwidth]{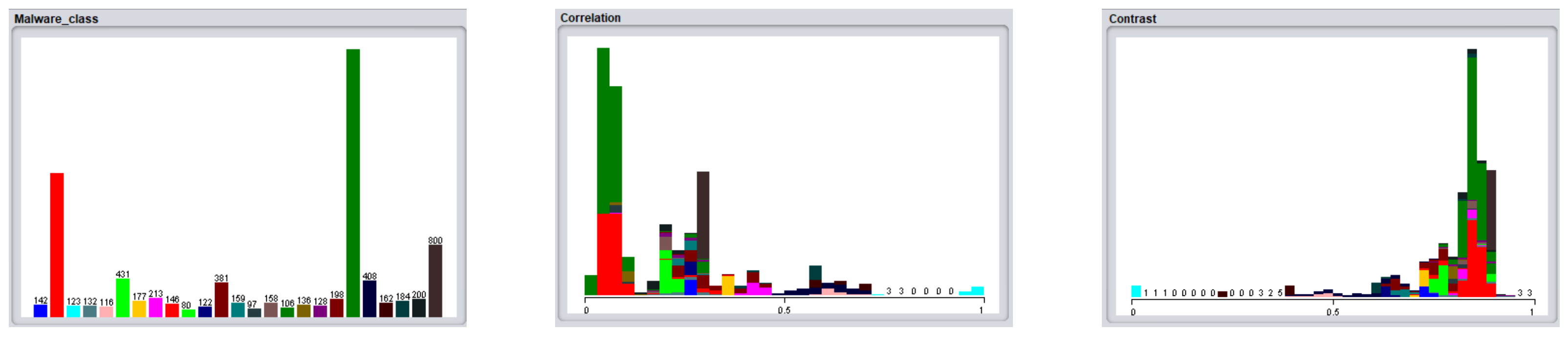}
    \vspace{1em}
    \caption{Visualization for 25-sub classes; (a) All malware sub-classes, (b) Correlation feature distribution and (c) Contrast feature distribution}
    \label{fig:visu_sub}
    \figSpace
\end{figure*}

Usually the dataset that is built using the extracted features is not `clean' and needs a lot of preprocessing to ensure optimal detection accuracy of the classifiers and its correct functioning as well. In our case, since we built the feature extraction and CSV scripts such that it was preprocessed even before the data was saved to CSV files hence eliminating the need of `cleaning' the dataset. The only thing that we needed to do was data normalization which is a prerequisite of the PCA analysis. We normalized the dataset with the help of a filter available in the Weka tool named ``normalize". This scales and normalizes and entire columns in the dataset and also does the same across the columns. This process was much needed to ensure our classification stage does not take any performance hit. Figures \ref{fig:visu_six} and \ref{fig:visu_sub} show the visualization of the main and the sub-classes of malware dataset for correlation and contrast features. It shows how the data is distributed across various classes of malware and we have chosen only those graphs that have distribution spread across the x-axis, whereas all the other feature graphs were not so much spread and the columns overlapped each other - this was resolved when we trained the ML classifiers with the dataset because the problem is solved in higher dimensional space as against 2D visualization. 

\subsubsection{Bag of Classifiers} After the visualization was done on the dataset, we have trained our classifier(s) with the dataset. We have named this section as 'bag of ML classifiers' because we have trained multiple classifiers -each of a different type- to observe and draw conclusions as to which one performs the best. I our case, we have used the supervised type of classifiers which in our case have shown promising results as against unsupervised methods that rendered less than 30\% detection accuracy and needed further improvisations and hence we did not include those results here. We used the Weka tool \cite{Hall_weka'09} to perform all the ML training and testing. We have used the following types of classifiers which are unique in themselves:

\subsubsection{{Naive Bayes}}
Naive Bayes is a simple, yet effective and commonly-used, machine learning classifier. It is a probabilistic classifier that makes classifications using the Maximum A Posteriori decision rule in a Bayesian setting. It can also be represented using a very simple Bayesian network. Naive Bayes classifiers have been especially popular for text classification, and are a traditional solution for problems such as spam detection \cite{naive_bayes}.

\subsubsection{{Random Forest}}
Random Forest is a flexible, easy to use machine learning algorithm that produces, even without hyper-parameter tuning, a great result most of the time. It is also one of the most used algorithms, because it’s simplicity and the fact that it can be used for both classification and regression tasks.Random Forest is a supervised learning algorithm. Random forest builds multiple decision trees and merges them together to get a more accurate and stable prediction \cite{random_forest}.

\subsubsection{{Logistic Regression}} 
Logistic regression is a classification algorithm used to assign observations to a discrete set of classes. Unlike linear regression which outputs continuous number values, logistic regression transforms its output using the logistic sigmoid function to return a probability value which can then be mapped to two or more discrete classes. LR could help use predict whether the class under consideration is a rootkit or a worm. Logistic regression predictions are discrete - only specific values or classes are allowed \cite{logistic_regression}. All the results of the classifiers are presented in Section \ref{results}.

\subsubsection{{Ontograph with Protege Tool}}

\begin{figure*}
    \figSpace
    \centering
    \includegraphics[width=1\textwidth]{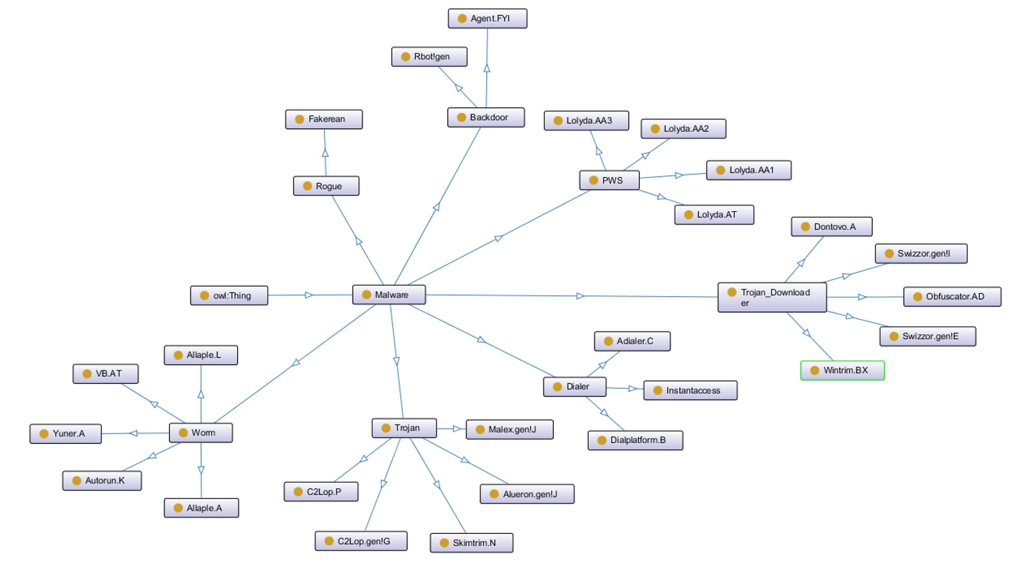}
    \vspace{10pt}
    \caption{Ontograph for malimg-dataset\vspace{10pt}}
    \vspace{1pt}
    \label{fig:ontograph}
    \figSpace
\end{figure*}

Figure \ref{fig:ontograph} illustrates the ontograph for malimg dataset. We obtained this ontograph by using protege tool. Protege tool is a free, open source ontology editor and a knowledge management system. Protege provides a graphic user interface to define ontologies. It also includes deductive classifiers to validate that models are consistent and to infer new information based on the analysis of an ontology \cite{protege}.

\subsection{Experimental Results}

\label{results}
We here discuss and present all the results that we gathered for various sections of our project discussed so far. The results are presented in the same order as we have mentioned in the earlier sections. 

\subsubsection{PCA Analysis} The results of the 6-main class and 25-sub class PCA analysis is as shown in Table \ref{tbl:sub_class_pca} respectively. The tables clearly indicate the features that were used to build the new principle components (PCs) and how much information each of them carries is mentioned in the `attribute' column of the tables. These PCs can be used to feed and train the ML-classifier exploiting the benefits mentioned previously. It is to be noted that we did not use the results of the PCA for training our classifiers because with the existing dataset we could get better accuracy and performance of the ML-models with limited time complexity and hence it was not needed to use PCA results but we still did include the results just to show that PCA is also a good option for achieving better accuracy with less time complexity. 


\begin{table*}[!h]
\tableSpace
\centering

\caption{PCA Results for 25-sub class of malware}
\vspace{1em}
\label{tbl:sub_class_pca}
\scalebox{1}{
\begin{tabular}{ll}
\hline
Ranked & Attributes \\ 
\hline
\hline 
0.7445 & 0.342Energy-0.341Homogeneity \\
\hline
0.6981 & 2 -0.567Malware\_class=Allaple.A-0.357Malware\_class=Lolyda.AA3 \\
\hline
0.6591 & 3 -0.776Malware\_class=Allaple.L+0.524Malware\_class=Allaple.A \\
\hline
0.6234 & 4 -0.681Malware\_class=Yuner.A-0.253Contrast \\

\hline
\hline

\end{tabular}}
\tableSpace
\end{table*}

\subsubsection{Classification Results}
\begin{figure*}
    \figSpace
    \centering
    \includegraphics[width=1\textwidth]{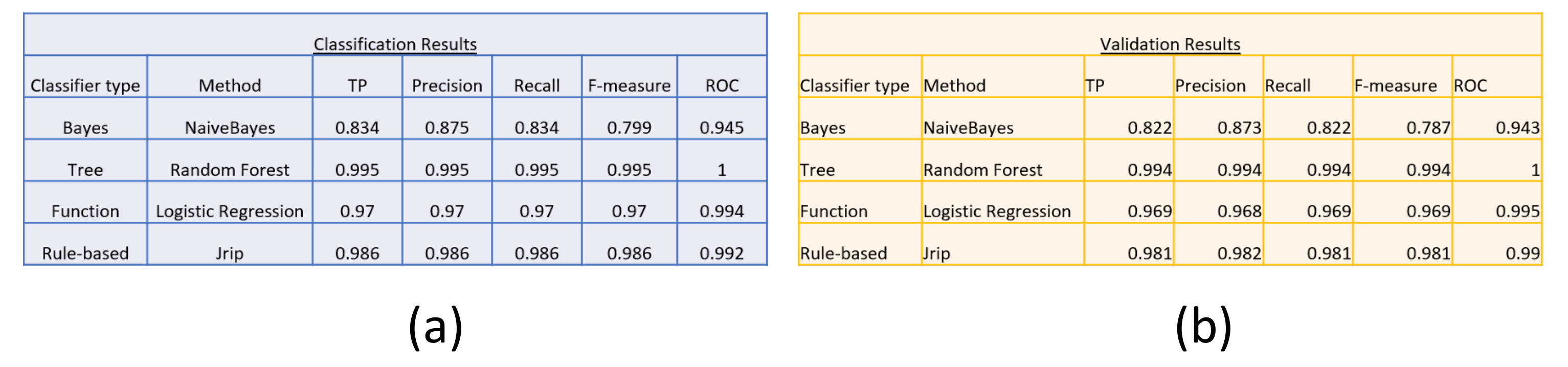}
    \vspace{1em}
    \caption{Results for 6-main classes (a) Classification results, (b) Validation Results}
    \label{fig:six_class_results}
    \figSpace
\end{figure*}

\begin{figure*}
    \figSpace
    \centering
    \includegraphics[width=1\textwidth]{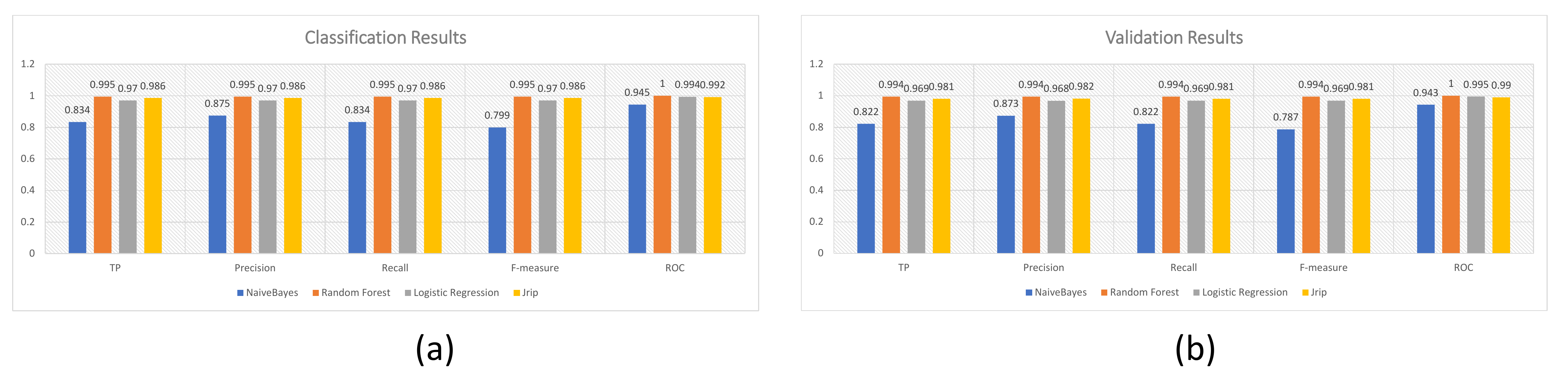}
    \vspace{1em}
    \caption{Results for 6-main classes (a) Classification bar graphs, (b) Validation bar graphs}
    \label{fig:six_class_plots}
    \figSpace
\end{figure*}

\begin{figure*}
    \figSpace
    \centering
    \includegraphics[width=0.95\textwidth]{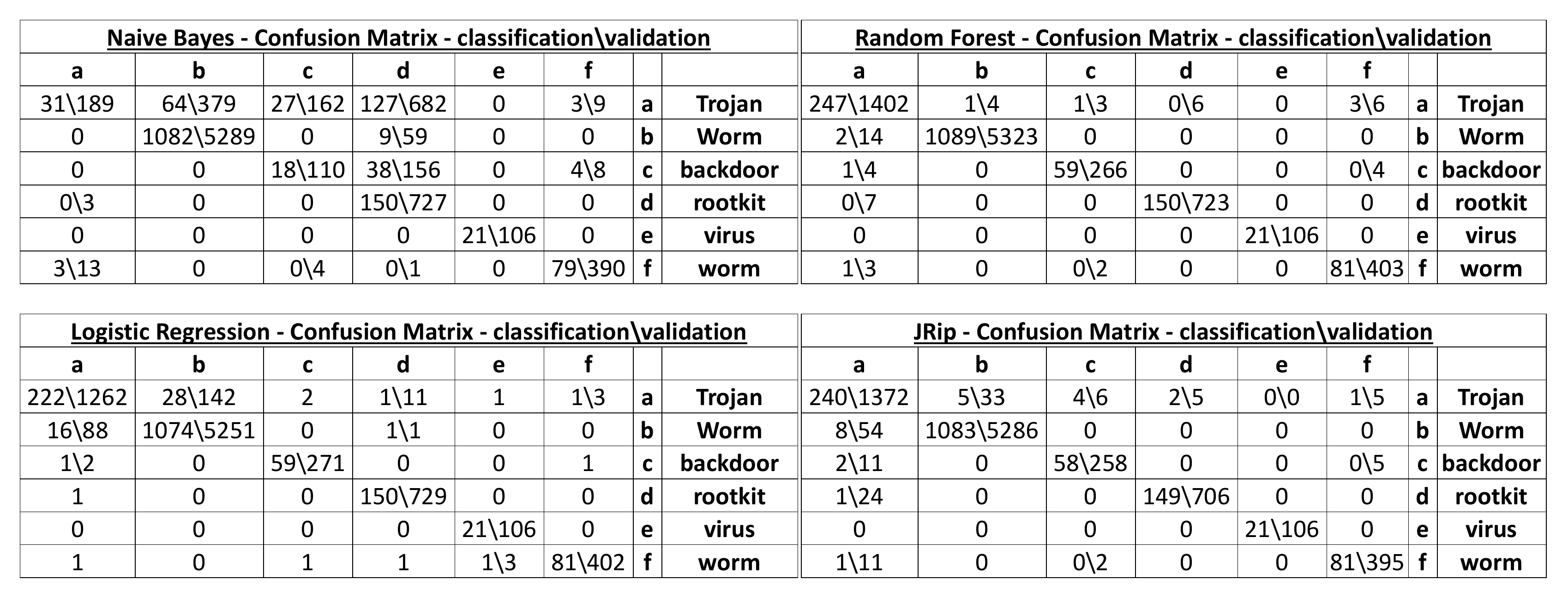}
    \vspace{1em}
    \caption{Confusion matrices for all the Classifiers for six main malware class}
    \label{fig:six_class_confusion}
    \figSpace
\end{figure*}

\begin{figure*}
    \figSpace
    \centering
    \includegraphics[width=1\textwidth]{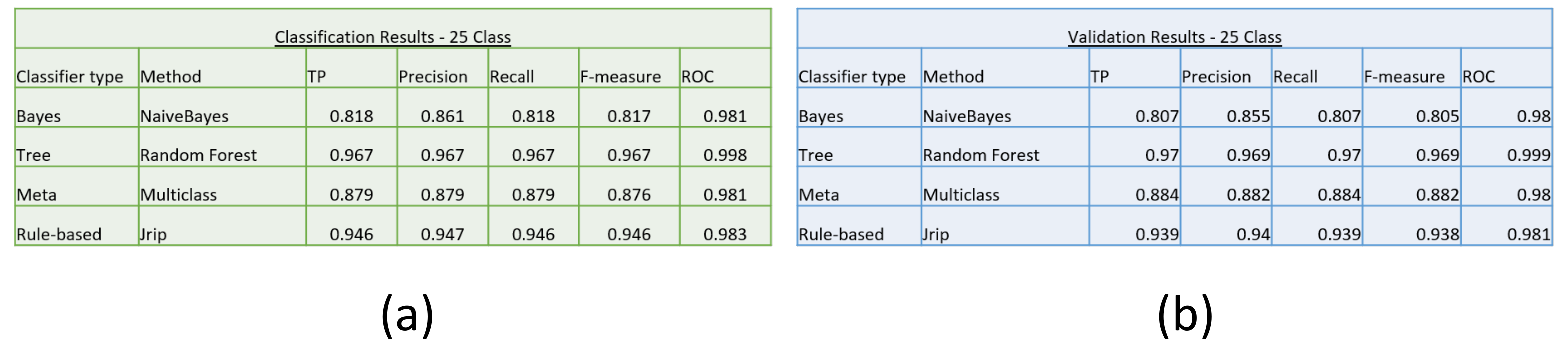}
    \vspace{1em}
    \caption{Results for 25-sub classes (a) Classification results, (b) Validation Results}
    \label{fig:two_five_results}
    \figSpace
\end{figure*}

\begin{figure*}
    \figSpace
    \centering
    \vspace{2em}
    \includegraphics[width=1\textwidth]{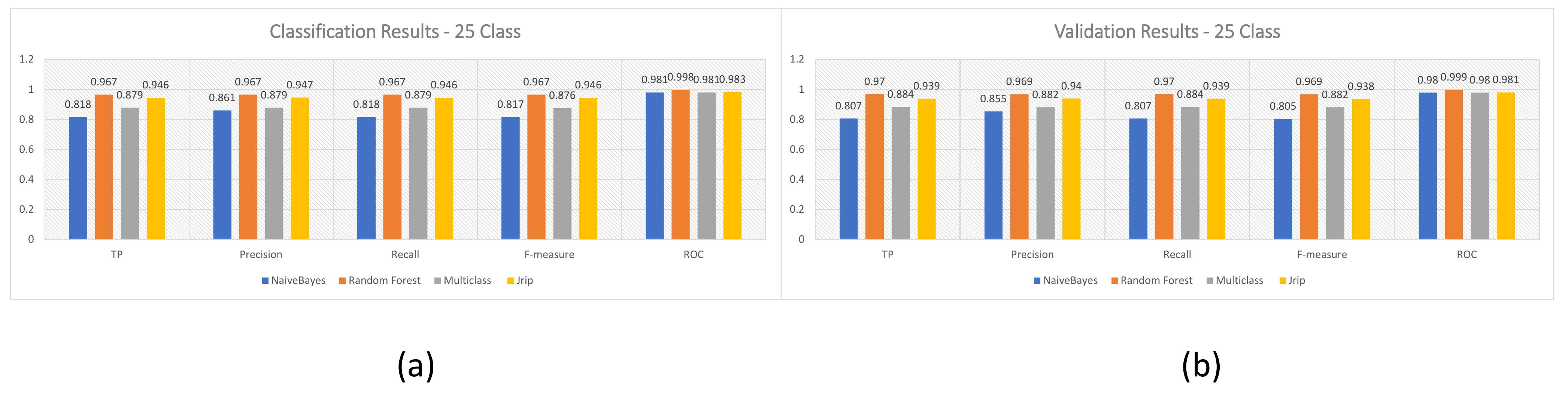}
    \vspace{1em}
    \caption{Results for 25-sub classes (a) Classification bar graphs, (b) Validation bar graphs}
    \label{fig:two_five_plots}
    \figSpace
\end{figure*}

The results of all the classifiers that we have used are presented here with the conclusions that we can draw from them. Figure \ref{fig:six_class_results}(a),(b) show classification accuracy and other performance metrics. Figure \ref{fig:six_class_results} shows the results of 6-main classes of malware after training and testing with 4-different machine learning models. We have used 80-20\% train/test while training the models. Figure \ref{fig:six_class_plots} shows the bar graph plots for the same data discussed earlier. The bar graphs show results with different performance metrics and we can conclude that all the classifiers used satisfy the given dataset and the problem of linear EDA graphs where all the plots were kind of linear were solved after classification results as the classification is done in higher dimensions. We did 2 fold validation on each classifier and did not go with higher fold validations as it took a lot of time to validate. The same results have been plotted for 25-sub classes of malware as shown in Figure \ref{fig:two_five_results} and Figure \ref{fig:two_five_plots}. These graphs show that with 25 class dataset, the results are better although some of the classes overlap which might be because two or more classes show similar characteristic features. 
Figure \ref{fig:six_class_confusion} shows the confusion matrix for 

\subsection{Related Works}

This section discusses some of the previous works that have been published in the past that have proposed methods to detect malwares. Some of the techniques discussed here are software-based while others are hardware-based.
Works in \cite{sayadi-CF18,DAC4'17} extensively described how hardware performance counters (HPCs) can be used to detect anomalies in applications and classify malware as against benign employing ML-classifiers. The authors have used HPCs to feed to different set of classifiers and presented their results. They have also proposed malware detection for resource constrained systems where performance counters are limited and where systems resources have to used sparingly. Hence, they propose the use of Ensemble learning methods to boost the performance of general ML-classifiers.
\cite{Demme'13} has discussed about the feasibility of using HPCs for malware detection. They have also used ML-models to classify applications and supported their claim. In \cite{Sanket_ICMLA'19, Sanket_ICTAI'19} authors detect stealthy malwares by converting malware binaries into grayscale images and then extracting patterns by performing raster scanning. The grayscale images are further represented as sequence of patterns which are further used for sequence classification using RNN-LSTM's. Work in \cite{Sanket_CASES'19} introduces a hybrid approach which utilizes the microarchitectural traces obtained through on-chip embedded hardware performance counters (HPCs) and the application binary for malware detection. The obtained HPCs are fed to multi-stage machine learning (ML) classifier for detecting and classifying the malware. Authors in \cite{Sanket_DAC'21}  presents a collaborative machine learning (ML)-based malware detection framework. It introduces a) performance-aware precision-scaled federated learning (FL) to minimize the communication overheads with minimal device-level computations; and (2) a Robust and Active Protection with Intelligent Defense strategy against malicious activity (RAPID) at the device and network-level due to malware and other cyber-attacks. 
\cite{Rootkit-Singh} has proposed how kernel-level rootkits can be detected using HPCs on hardware level. They have described the process of training ML-models with the acquired HPCs and then presented results in support of their claim. They have tested their proposed mechanism by feeding the detector with both rootkit and clean traces. 
Authors in \cite{Bahador'14} interestingly used SVD (Singular-value decomposition) matrix in collaboration with HPCs to train ML-models to detect malwares. This is kind of partial software and partial hardware based approach in detecting anomalies. 
\cite{Jacob'08} has proposed a software based approach to detect malicious piece of code.

%% file: ChapterThree.tex
\section{Need for RRAM-Neuromorphic Architecture based Defense Against Adversarial Attacks}

As, described in the introduction, the adversarial attacks can be broadly classified into two categories: (a) poisoning attacks and (b) evasion attacks. Poisoning attacks are attacks on the ML classifier during the training phase \cite{Nelson2008ExploitingML, rubinstein2009antidote, biggio2012poisoning, munoz2017towards, sshukla4_rram}, and the evasive attacks are targeted for inference stage of machine learning techniques. Poisoning attacks are appropriate for online environments because they focus on attacking the classifiers during the training phase. 
Therefore, in this work we focus on the evasive attacks, as many of the existing machine learning works are primarily offline learning based and are constrained by resources and the computational time requirements.

Rise in the types of adversarial attacks led to development of adversarial defense techniques. Some of the prominent software based adversarial defenses includes adversarial training \cite{szegedy2013intriguing, goodfellow2014explaining}, defensive distillation \cite{Papernot2016DistillationAA} and MagNet \cite{meng2017magnet}. Even though these adversarial defense show some robustness against adversarial samples, they also have major drawbacks and weaknesses. More details over the adversarial defenses is presented in Section \ref{attacks and defense}. 

As the aforementioned defense is developed based on software, researchers have started to shift the focus from software to hardware \cite{Abhijitt_8,Abhijitt_9,Abhijitt_10,Abhijitt_12,Abhijitt_13, Abhijitt_14}. There have been a new research trending which comprises of leveraging neuromorphic computing to provide a robust defense against the adversaries. In \cite{bhattacharjee2020rethinking} authors demonstrate the advantage conferred by the non-idealities present in analog crossbars in terms of adversarial robustness. Authors in the paper \cite{kim2019neuromorphic} propose a neuromorphic approach based on sparse coding. These neuromorphic based techniques are designed specifically for a targeted attack such as one-pixel attack and FGSM attack. The interoperability of these defense against other pool of adversarial attacks still remains a concern. 

In this work, we first provide an overview of evasive attacks on the ML classifiers. Further, we present different existing defense techniques for the adversarial attacks. As FGSM is one of the fastest evasive attacks, an in-depth discussion regarding the FGSM adversarial attack is provided. In this work, we look at initially introduced
defense against adversarial samples, Adversarial training is one of the defense techniques introduced for adversarial attacks. Further, in this paper we explore the potency of adversarial attacks on a reconfigurable RRAM-based Neuromorphic Architecture. We also propose a method for mitigating adversarial attacks in deployed IoT devices combining precise software training algorithms and the reconfigurable RRAM-based Neuromorphic architecture.

\section{Background}
\label{attacks and defense}
Adversarial samples are generated by introducing crafted perturbations into the normal input data generated.  This makes the adversarial data look similar to the normal input data, but still the machine learning model mispredicts the class with a high probability. These adversarial samples can be considered as an optical illusion for the ML classifiers. In this section, we present different techniques widely used for generating the adversarial samples, and review some of the popular defense techniques deployed. 

\subsection{Adversarial Attacks}

Here we present an overview of some of the adversarial attacks that are effective against machine learning classifiers. 

\subsubsection{Fast Gradient Sign Method (FGSM)}
The most common technique to perform adversarial attack is to perturb the image with gradient of the loss with respect to the image or
input. Then gradually increase the magnitude of the perturbation until the image is misclassified.

Fast Gradient Sign method (FGSM) \cite{goodfellow2014explaining} is one of the first known adversarial attacks. The complexity to generate FGSM attack is
lower compared to other adversarial attacks, even against deep
learning models. This technique features
low complexity and fast implementation. Consider a ML classifier
model with $\theta$ as the parameter, $x$ being the input to the model,
and $y$ is the output for a given input $x$, and L($\theta$, $x$,$y$) be the cost
function used to train the neural network. Then the perturbation
with FGSM is computed as the sign of the model’s cost function
gradient. The adversarial perturbation generated with FGSM \cite{goodfellow2014explaining}
is mathematically given as

\begin{equation}
    x^{adv} = x + \epsilon sign(\nabla_{x} L(\theta, x, y))
\end{equation}

where $\epsilon$ is a scaling constant ranging between 0.0 to 1.0 is set
to be very small such that the variation in $x$ ($\delta x$) is undetectable.
One can observe that in FGSM the input $x$ is perturbed along each
dimension in the direction of gradient by a perturbation magnitude of $\epsilon$. While, a small $\epsilon$ leads to well-disguised adversarial samples, a large $\epsilon$, is likely to introduce large perturbations.

\subsubsection{Basic Iterative Method (BIM)}

From previous discussion it can be observed that, FGSM adds perturbation in each of the dimension, however, no optimization on perturbations are performed. In \cite{Kurakin2017AdversarialEI}
Kurakin proposed an iterative version of FGSM, called as Basic iterative method (BIM). 
BIM is an extension of FGSM technique, where instead of applying the adversarial perturbation once with $\epsilon$, the perturbation is applied multiple times iteratively with small $\epsilon$. This produces a recursive noise on the input and optimized application of noise. It can be mathematically represented as follows:

\begin{equation}
\begin{split}
    x_{0}^{adv} = x\\ 
    x_{N+1}^{adv} = Clip_{x, \epsilon}(x^{adv}_{N} + \epsilon sign(\nabla_{x} L(\theta, x^{adv}_{N}, y)))
\end{split}
\end{equation}

In the above mathematical expression,$Clip_{x, \epsilon}$ represents the clipping of the adversarial input magnitudes such that they are within the neighborhood
of the original sample $x$. This technique allows more freedom for
the attack compared to the FSGM method because the perturbation can
be controlled and the distance of the adversarial sample from the
classification boundary can be carefully fine-tuned. The experimental results presented in \cite{Kurakin2017AdversarialEI} have shown that BIM can cause higher misclassifications
compared to the FGSM attack on the Imagenet samples.

\subsubsection{Momentum Iterative Method (MIM)}

The momentum method is an accelerated gradient descent technique that accumulates the velocity vector in the direction of the
gradient of the loss function across multiple iterations. In this technique, the previous gradients are stored, which aids in navigating
through narrow valleys of the model, and alleviate problems of
getting stuck at local minima or maxima. This momentum method
also shows its effectiveness in stochastic gradient descent (SGD) to
stabilize the updates. This MIM principle is applied in \cite{Dong2018BoostingAA} to generate adversarial samples. MIM has shown a better transferability
and shown to be effective compared to FGSM attack.

\subsubsection{DeepFool Attack}

DeepFool (DF) is an untargeted adversarial attack optimized for
$L_{2}$ norm, introduced in \cite{MoosaviDezfooli2016DeepFoolAS}. DF is efficient and produces adversarial
samples which are more similar to the original inputs as compared
to the discussed adversarial samples generated by FGSM and BIM
attacks. The principle of the Deepfool attack is to assume neural
networks as completely linear with a hyper-plane separating each
class from another. Based on this assumption, an optimal solution
to this simplified problem is derived to construct adversarial samples. As the neural networks are non-linear in reality, the same
process is repeated considering the non-linearity into the model.
This process is repeated multiple times for creating the adversaries.
This process is terminated when an adversarial sample is found i.e.,
misclassification happens. Considering the brevity and focus of the
current work, we limit the details in this draft. However, the interested readers can refer to the work in \cite{MoosaviDezfooli2016DeepFoolAS} for exact formulation of
DF.

\subsection{Adversarial Defenses}

So far, the different adversarial attack techniques are discussed. Here,
we discuss some of the prominent existing defenses against the above discussed attacks.

\subsubsection{Adversarial Training}
Adversarial training is one of the preliminary solutions for making the ML classifiers robust against the adversarial examples, proposed in \cite{shaham2015understanding}. The idea is to train the ML classifier with the adversarial examples so that the ML classifier can have adversarial information \cite{goodfellow2014explaining, szegedy2013intriguing} and adapt its model based on the learned adversarial data. However, one of the major drawbacks of this technique is to anticipate the type of attack and train the classifier based on those attacks and determining the criticality of the adversarial component.

\subsubsection{Defensive Distillation}
Defensive distillation is another defense technique proposed in \cite{Papernot2016DistillationAA}. The idea is to train the classifier using the distillation training techniques and hides the gradient between softmax layer and the presoftmax layer. This makes it complex to generate adversarial examples directly on the network \cite{Carlini2017TowardsET}, as the knowledge is imparted from a bigger network during the training process. However, \cite{hinton2015distilling} shows that such a defense can be bypassed with one of the following three strategies: (1) choosing a more proper loss function; (2) calculation of gradients from pre-softmax layer rather than softmax layer; or (3) attack an easy-to-attack dummy network first and then transfer to the distilled network, similar to the distillation
defense. The generation of adversaries can be simpler if the attacker knows the parameters and architecture of the defense network i.e., whitebox attack.

\subsubsection{MagNet}
MagNet is proposed in \cite{meng2017magnet}, where a two-level strategy with detector and reformer is proposed. In the detector phase(s), the system learns to differentiate between normal and adversarial examples by approximating the manifold of the normal examples. This is performed with the aid of auto-encoders. Further, in the reformer, the adversarial samples are moved close the manifold of normal samples with small perturbations. Further using the diversity metric, the MagNet can differentiate the normal and adversarial
samples. MagNet is evaluated against different adversarial attacks presented previously and has shown to be robust in \cite{meng2017magnet}.

\subsubsection{Detecting Adversaries}
Another defense technique is to detect adversarial examples with the aid of statistical features \cite{Grosse2017OnT} or separate classification networks \cite{Metzen2017OnDA}. In \cite{Metzen2017OnDA}, for each adversarial technique, a DNN classifier is built to classify whether the input is a normal sample or an adversary. The detector was directly trained on both normal and adversarial examples. The detector shows good performance when the training and testing attack examples were generated from the same process and the perturbation is large enough. However, it does not generalize well across different attack parameters and attack generation processes.

\section{Proposed RRAM-based Neuromorphic Architecture}

\begin{figure}[h]
    \figSpace
    \centering
    \includegraphics[width=0.85\textwidth]{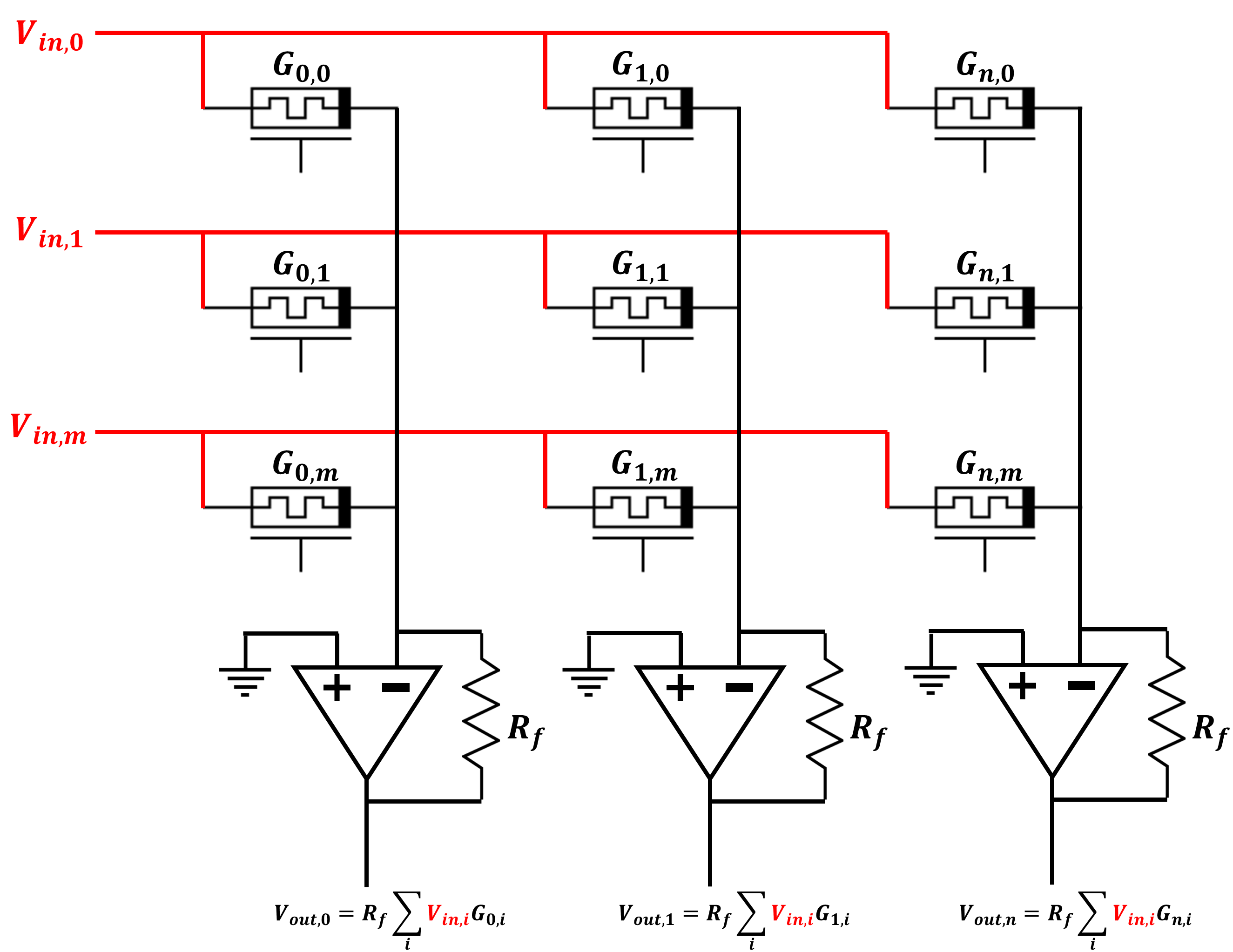}
    \caption{Gated-RRAM neuromorphic crossbar performing MAC operation on voltage-encoded input vector and conductance-encoded weight matrix}
    \label{fig:rram_crossbar}
    \figSpace
\end{figure}

The Internet of Things (IoT) continues to expand and there is increasing interest in computing at the edge of the network especially in machine learning applications. 
Many current implementations of machine learning, and more specifically Deep Neural Networks (DNNs), have large power and computational resource requirements.
The high memory bandwidth and power requirement prevent implementation in low-power real-time applications.
Neuromorphic architectures have been explored for near-memory and in-memory computing to for low-power high memory bandwidth implementation of neural networks. Particularly, two-terminal RRAM devices in crossbar arrays have been extensively investigated to store weight matrix and perform in-memory computing.  The RRAM-based weight matrix crossbar of neural networks can be visualized in Figure \ref{fig:rram_crossbar}. The input vector gets multiplied by the weight matrix to produce the composition of the neurons at the output. A winner-take-all or softmax activation function can then be used to determine the winning neuron and thus the classification. In spite of  tremendous progress in this area two-terminal RRAM devices suffers from issues such as high write current, convoluted read and write strategies, and need for a selector diode to mitigate sneak currents. Recently reported gated-RRAM devices offers to solve these issues and provide opportunities for simultaneous read and write which will be very beneficial for adjusting the weights as needed. 

\begin{figure}[h]
    \figSpace
    \centering
    \includegraphics[width=0.85\textwidth]{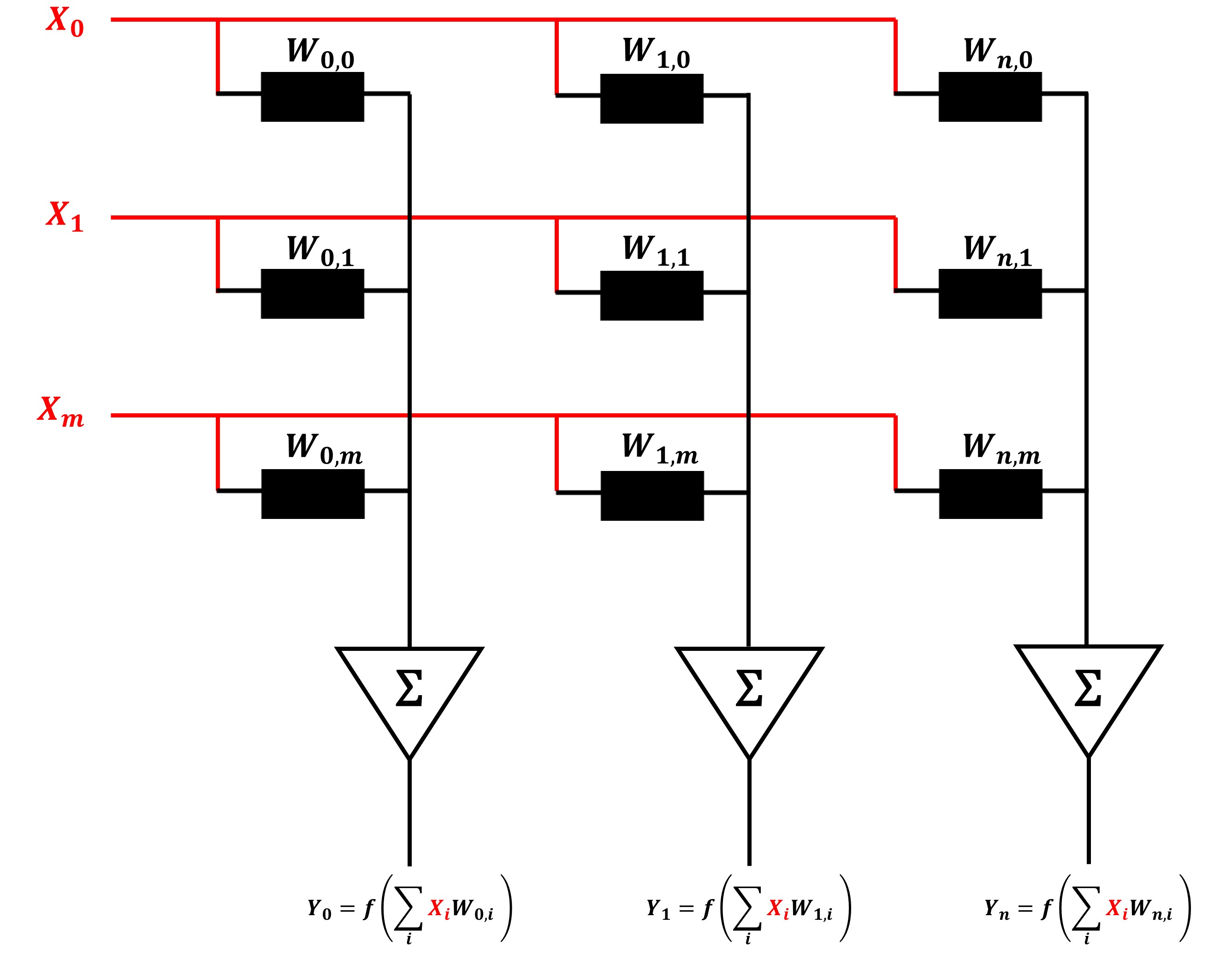}
    \caption{Feed-forward neural network crossbar performing MAC operation on input vector and weight matrix}
    \label{fig:crossbar}
    \figSpace
\end{figure}

\par Gated-RRAM have been investigated as memristive synaptic devices for in-memory computing of the multiply-accumulate (MAC) behavior of neurons.
Additionally, integrating multi-state gated-RRAM allows for a more dense memory crossbar since each gated-RRAM device can store multiple bits.
The weight matrix crossbar of neural networks can be visualized in Figure \ref{fig:crossbar}. 
The input vector gets multiplied by the weight matrix to produce the composition of the neurons at the output.
A winner-take-all or softmax activation function can then be used to determine the winning neuron and thus the classification.
Similarly in a neuromorphic gated-RRAM crossbar, the product of the input voltage and gated-RRAM conductance produces a current at each synaptic branch of a neuron.
The summation amplifier at the output then computes the multiply-accumulate function of the input vector, applied as voltages and the weight vector of gated-RRAM conductances of the neuron as shown in Figure \ref{fig:rram_crossbar}.


\begin{figure}
    \figSpace
    \centering
    \includegraphics[width=0.5\textwidth]{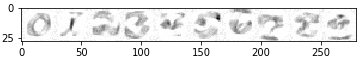}
    \caption{Weight map of keras-trained weights on MNIST}
    \label{fig:weight_maps}
    \figSpace
\end{figure}

\begin{figure}
    \figSpace
    \centering
    \includegraphics[width=0.5\textwidth]{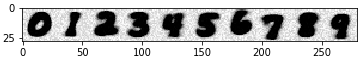}
    \caption{Weight map of STDP-based training on MNIST}
    \label{fig:weight_maps}
    \figSpace
\end{figure}

\par The reconfigurability of gated-RRAM allows for synaptic weights to be programmed into the neuromorphic crossbar. 
The gated-RRAM device can be potentiated or depressed by applying a positive or negative bias at the gate respectively.
We were able to train weights using the keras package in Python 3.9 on the MNIST dataset and then write them to the neuromorphic crossbar.
However, the conductance curve of the gated-RRAM devices influences how the off-chip trained weights are mapped to the RRAM conductance space.
This may effect the behavior of the neuromorphic-implemented model depending on the non-linearity of the conductance curve of the RRAM devices.
Additionally, on-chip training directly on the RRAM crossbar has been explored and has demonstrated competitive training accuracy \cite{STLT}.
In Bailey et al. \cite{STLT}, a spike-timing dependent plasticity-based (STDP) algorithm is used to tune the weights of neuromorphic architecture.
This training algorithm results in different weight map, thus a different trained model, than the keras-trained weights as shown in Figure \ref{fig:weight_maps}.

\begin{figure}
    \figSpace
    \centering
    \includegraphics[width=0.5\textwidth]{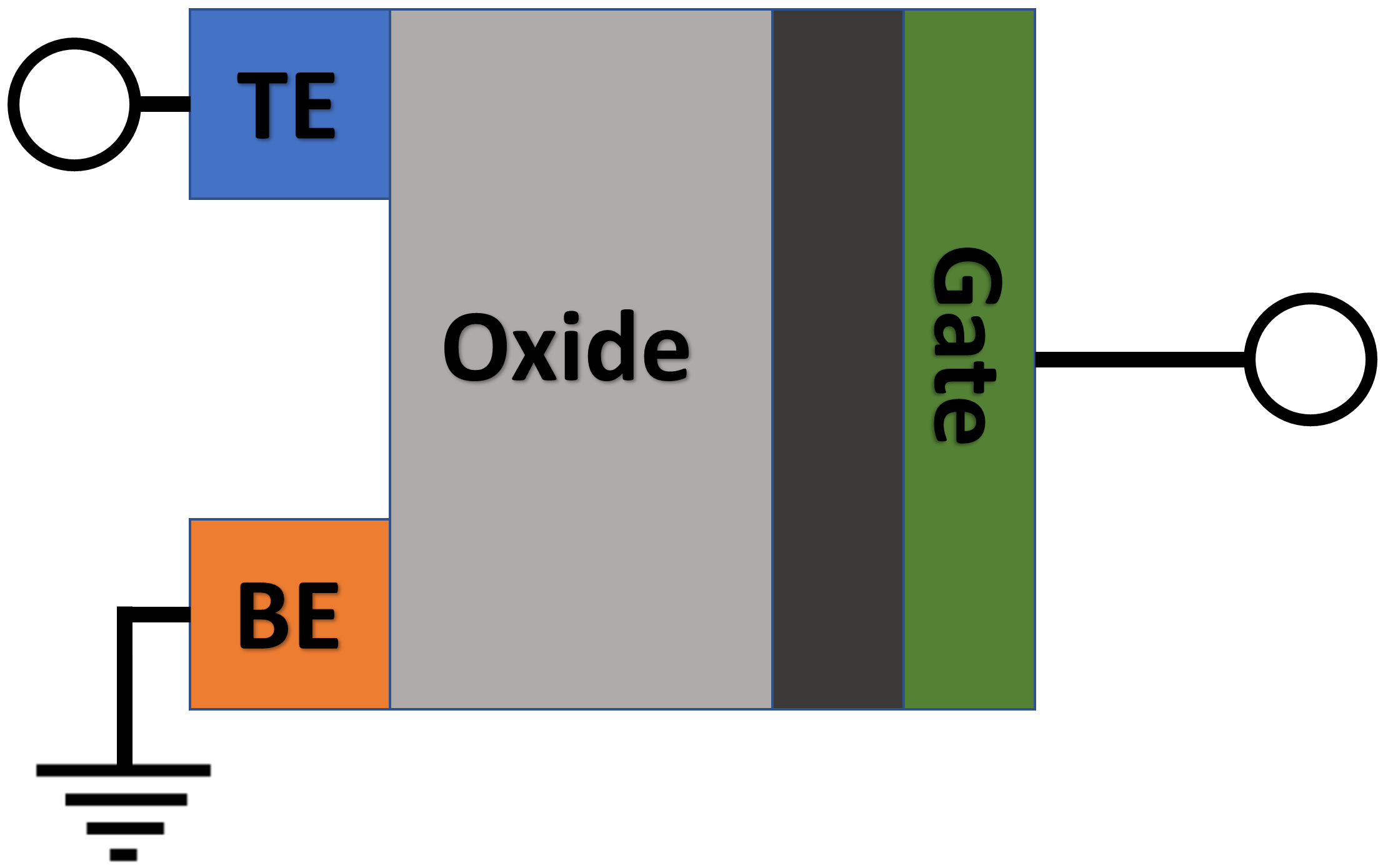}
    \caption{Gated-RRAM device model with top electrode (TE), bottom electrode (BE), gate, and channel oxide}
    \label{fig:RRAM}
    \figSpace
\end{figure}

\begin{figure}
    \figSpace
    \centering
    \includegraphics[width=0.75\textwidth]{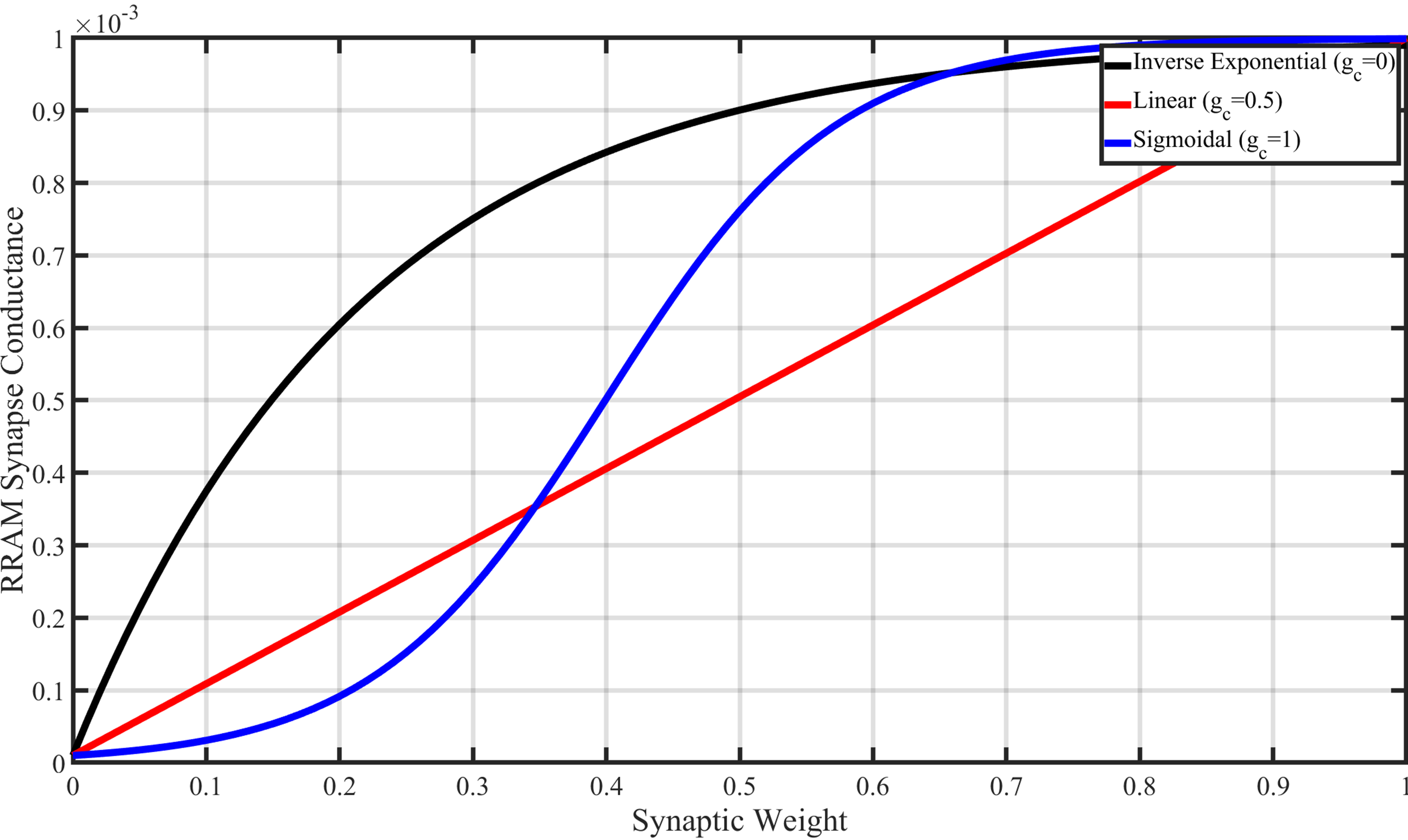}
    \caption{Conductance curves of gated-RRAM tuned through gated-synaptic model in \cite{GSD}}
    \label{fig:RRAM_cond}
    \figSpace
\end{figure}

Gated-resistive random-access memory (gated-RRAM) has been recently report and investigated as a gated-synaptic device for neuromorphic architectures \cite{3T-RRAM}.
Gated-RRAM is a memristive device consisting of a top and bottom electrode connected to an oxide channel as shown in Figure \ref{fig:RRAM}.
A gate terminal is coupled to the channel oxide through an insulating layer allowing for a bias to be applied across the channel oxide while simultaneously applying a bias across the top and bottom electrode.
This behavior allows for simultaneous reading and writing the gated-RRAM device and allows minimal programming circuitry.
The gated-RRAM device can be potentiated by applying a positive bias to the gate to cause the oxygen defects to drift toward the top and bottom electrode increasing the conductance of the channel.
The device can then be depressed by applying a negative bias to the gate to cause the oxygen defects to diffuse back towards the channel oxide lowering the conductance of the channel.
The conductance states or curve a device follows as it is potentiated and depressed is material dependent.
The conductance of these devices has been shown to follow sigmoidal, linear, or inverse exponential trends as shown in Figure \ref{fig:RRAM_cond} \cite{GSD}.
These conductance curves can determine the distribution of the conductance states of the gated-RRAM device and are controlled by $g_{c}$ in the gated-synaptic device model \cite{GSD}.
Similar to the two-terminal RRAM crossbar arrays,  in a neuromorphic gated-RRAM
crossbar, the product of the input voltage and gated-RRAM conductance
produces a current at each synaptic branch of a neuron.
The summation amplifier at the output then computes the multiplyaccumulate
function of the input vector, applied as voltages and
the weight vector of gated-RRAM conductances of the neuron as
shown in Figure \ref{fig:rram_crossbar}.

\section{Results}

In this section, we evaluate and present the performance on the MNIST digit dataset \cite{mnist}. The adversarial attacks are generated using Cleverhans library \cite{cleverhans}.

Table \ref{tbl:adv_results} reports the performance of the employed neural network
on MNIST Digits dataset.
Normal Classification Accuracy: In the absence of adversarial samples, the classifier achieves an accuracy of 98.25\%, loss of 0.088, precision of 0.98, and recall of 0.98.
We also report the accuracy of neural network in the presence of adversarial attacks in Table \ref{tbl:adv_results}. The number of adversarial samples are 10,000 in
each case, and one can observe that in the presence of adversaries
the classification accuracy falls to as low as 1.13\%. With the increase
in $\epsilon$, the accuracy decreases in case of FGSM, MIM and BIM.

\begin{table}[]
\tableSpace
\caption{Accuracy of neural network before and after adversarial attack}
 \vspace{1em}
\centering
\scalebox{1} {

\begin{tabular}{|c|c|c|c|c|}
\hline
Adversarial attack & Parameter & No Attack & With Attack \\
\hline
FGSM               &  $\epsilon$ = 0.3       & 98.25 \%    & 8.54        \\
\hline
BIM                & $\epsilon$ = 0.3       & 98.25 \%     & 1.34        \\
\hline
MIM                & $\epsilon$ = 0.3       & 98.25 \%    & 1.28        \\
\hline
DF                 & MI = 50     & 98.25 \%    & 1.13
\\
\hline
\end{tabular}
}
\label{tbl:adv_results}
\tableSpace
\end{table}

We present the effect of adversarial training by training neural network with adversarial data generated using different adversarial attacks and present the accuracy results in Table \ref{tbl:adv_train_results}. The results show that after performing adversarial training, the accuracy of the neural network to classify data improves as reported in the Table \ref{tbl:adv_train_results}. However, the accuracy can further degrade if the attack parameters ($\epsilon$ in this case) are modified.

\begin{table}[h]
\tableSpace
\caption{Accuracy (\%) of MNIST Classification under Different Adversarial Attacks on Different Adversarial Trained Networks}
 \vspace{1em}
\centering

\begin{tabular}{|c|c|c|c|c|c|}
\hline
Attack & BIM & MIM &  FGSM & DF \\                      
\hline
Parameter & $\epsilon=0.3$ & $\epsilon=0.3$ & $\epsilon=0.3$ & MI=50 \\
\hline
FGSM & 79.1 & 78.2 &  & 77.1 \\
\hline
DF      & 53.40 & 54.25 & 41.04 & \\
\hline
MIM  & 78.2 & & 73.7 & 77.1 \\
\hline
BIM   &  & 78.1 & 69.3 & 78.2 \\
\hline
\end{tabular}
\label{tbl:adv_train_results}
\tableSpace
\end{table}

\begin{table}[h]
    \tableSpace
    \caption{Accuracy (\%) of MNIST classification of RRAM neuromorphic architecture on MNIST dataset and different adversarial attacks. The conductance curve of the gated-RRAM devices is linear and there is no write variance present.}
     \vspace{1em}
    \centering
    \begin{tabular}{|c|c|c|c|c|c|}
    \hline
    \multicolumn{1}{|c|}{} & \multicolumn{5}{c|}{Adversarial Attack} \\
    \hline
    Training & No Attack & FGSM & BIM & DF & MIM \\
    \hline
    Keras & $90.44\%$ & $8.74\%$ & $8.95\%$ & $9.60\%$ & $8.91\%$ \\
    \hline
    STDP & $71.22\%$ & $13.09\%$ & $11.98\%$ & $11.69\%$ & $12.15\%$ \\
    \hline
    \end{tabular}

    \label{tab:adv_accuracy}
    \tableSpace
\end{table}

\begin{table}[h]
    \tableSpace
    \caption{Accuracy (\%) of MNIST classification of RRAM neuromorphic architecture, with varying conductance distributions, on MNIST dataset and different adversarial attacks. Additionally, there is no write variance present.}
     \vspace{1em}
    \centering
    \scalebox{0.9}{
    \begin{tabular}{|c|c|c|c|c|c|c|}
    \hline
    \multicolumn{1}{|c|}{} & \multicolumn{1}{|c|}{}& \multicolumn{5}{c|}{Adversarial Attack} \\
    \hline
    Training & RRAM $g_{c}$ & No Attack & FGSM & BIM & DF & MIM \\
    \hline
    Keras & $0.0$ & $85.53\%$ & $9.54\%$ & $9.63\%$ & $10.43\%$ & $9.59\%$ \\
    \hline
    Keras & $0.5$ & $90.44\%$ & $8.74\%$ & $8.95\%$ & $9.60\%$ & $8.91\%$ \\
    \hline
    Keras & $1.0$ & $65.21\%$ & $4.96\%$ & $5.12\%$ & $5.77\%$ & $4.97\%$ \\
    \hline
     STDP & $0.0$ & $72.29\%$ & $9.67\%$ & $9.08\%$ & $7.62\%$ & $9.15\%$ \\
    \hline
    STDP & $0.5$ & $71.22\%$ & $13.09\%$ & $11.98\%$ & $11.69\%$ & $12.15\%$ \\
    \hline
    STDP & $1.0$ & $72.00\%$ & $11.36\%$ & $10.44\%$ & $9.64\%$ & $10.50\%$ \\
    \hline
    \end{tabular}
    }
    
    \label{tab:cond_accuracy}
    \tableSpace
\end{table}

\begin{table}[]
    \tableSpace
    \caption{Accuracy (\%) of MNIST classification of RRAM neuromorphic architecture, with varying device variance, on MNIST dataset and different adversarial attacks. Additionally, the RRAM devices have a linear conductance distribution.}
    \vspace{1em}
    \centering
    
    \scalebox{0.82}
    {
    \begin{tabular}{|c|c|c|c|c|c|c|}
    \hline
    \multicolumn{1}{|c|}{} & \multicolumn{1}{|c|}{}& \multicolumn{5}{c|}{Adversarial Attack} \\
    \hline
    Training & RRAM $\sigma^{2}$ & No Attack & FGSM & BIM & DF & MIM \\
    \hline
    Keras & $0$ & $90.44\%$ & $8.74\%$ & $8.95\%$ & $9.60\%$ & $8.91\%$ \\
    \hline
    Keras & $10^{-6}$ & $90.42\%$ & $8.75\%$ & $8.96\%$ & $9.60\%$ & $8.92\%$ \\
    \hline
    Keras & $10^{-5}$ & $90.38\%$ & $8.84\%$ & $9.00\%$ & $9.64\%$ & $8.99\%$ \\
    \hline
    Keras & $10^{-4}$ & $90.41\%$ & $8.80\%$ & $9.09\%$ & $9.70\%$ & $9.01\%$ \\
    \hline
    STDP & $0$ & $71.22\%$ & $13.09\%$ & $11.98\%$ & $11.69\%$ & $12.15\%$ \\
    \hline
    STDP & $10^{-6}$ & $71.2\%$ & $13.08\%$ & $11.98\%$ & $11.68\%$ & $12.16\%$ \\
    \hline
    STDP & $10^{-5}$ & $71.19\%$ & $13.10\%$ & $12.00\%$ & $11.70\%$ & $12.20\%$ \\
    \hline
    STDP & $10^{-4}$ & $70.74\%$ & $13.26\%$ & $12.09\%$ & $11.75\%$ & $12.27\%$ \\
    \hline
    \end{tabular}
    }

    \label{tab:var_accuracy}
    \tableSpace
\end{table}

We then loaded the keras-trained weights to the RRAM neuromorphic architecture and tested it on the MNIST dataset and the various adversarial attacks.
Additionally, we also tested the performance of the on-chip STDP-trained weights on the same datasets.
Table \ref{tab:adv_accuracy} report the accuracy of both the keras-trained and STDP-trained weights with RRAM devices with completely linear conductance curve and no write variability.
In the absence of adversarial samples, the off-chip keras-trained network outperforms the on-chip STDP-trained weights but both training methods yield above 70\% accuracy.
However, with the introduction of adversarial samples, the accuracy is suddenly reduced to below 10\% for keras-trained weights and below 20\% for STDP-trained weights.
Additionally, we tested the two networks with varying conductance curves, as shown in Table \ref{tab:cond_accuracy}, and varying device write variability, as shown in Table \ref{tab:var_accuracy}.
In all cases, the accuracy of the RRAM architecture was reduced to below 20\% once adversarial samples were introduced.

\section{Proposed Method of Mitigation and Future Work}
\begin{figure}[h]
    \figSpace
    \centering
    \includegraphics[width=0.85\textwidth]{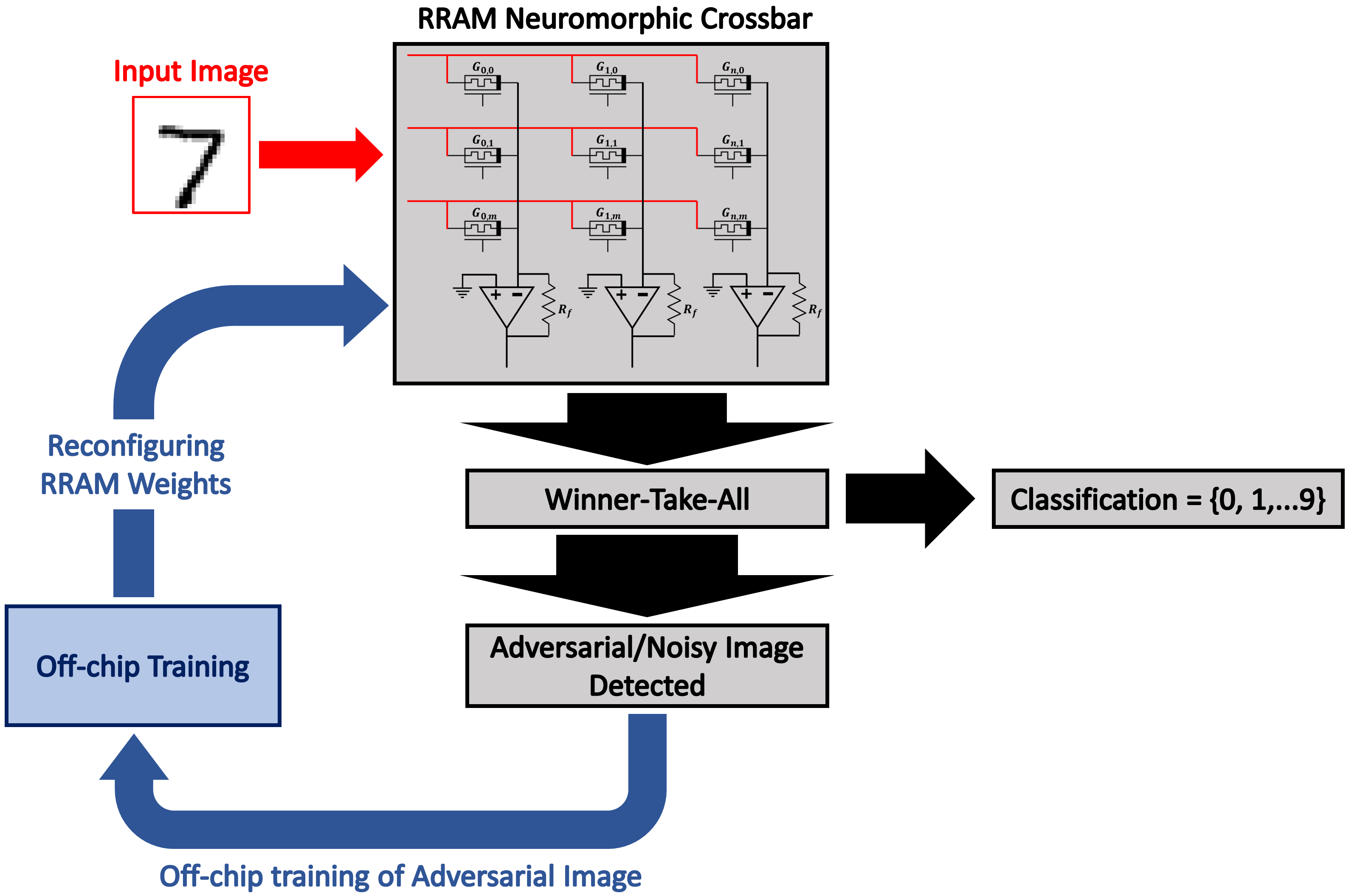}
    \caption{Integrated architecture of neuromorphic architecture with adversarial attack detection and mitigation}
    \label{fig:adv_arch}
    \figSpace
\end{figure}

In this work, we demonstrated that the gated-RRAM neuromorphic architecture serves as a versatile implentation of a hardware neural network.
The weights can be trained off-chip using software models and programmed during run-time.
Additionally, more biologically inspired algorithms can be applied to train the synaptic devices on-chip.
The gated-RRAM architecture is able to tune is learned model through different training algorithms and through different device characteristic such as conductance distribution and noise from device write variance.
However, we observed that this versatile model is still susceptible to an unknown adversarial attack.
\par
We would like to investigate an integrated system as shown in Figure \ref{fig:adv_arch}.
Our results in \ref{tbl:adv_train_results} demonstrate the ability of software networks to train on adversarial attacks for future mitigation of adversarial attacks.
We have also demonstrated the ability to load (write) pre-trained weights to the RRAM neuromorphic architecture model.
Therefore, a software network trained on adversarial models can be implemented using the RRAM neurormorphic architectures allowing it to be deployed in low-power real-time applications while mitigating attacks from adversaries.
Additionally, as shown in \ref{fig:adv_arch}, we would like to explore methods for the chip to detect unknown adversarial attacks so that it may later train on them and identify adversarial inputs or sources.
Our goal is to develop a reconfigurable architecture that is robust to adversarial attacks and with the capabilities to be integrated in low-power IoT devices or applications.

%% file: conclusion.tex
\section{Conclusion}

We have first described the process of converting malware binaries to images and then explained the process of feature extraction. With the help of our results we have supported our claim that malware binary images can be used to detect classes of malware from each other with high accuracy and with a variety of classifiers. We then have presented the confusion matrices to show the classification efficiency of the employed models. Protege tool has been used to obtain ontograph for the dataset based on the main and sub classes. We hope to develop this method and take it further ahead in classifying malwares from benign and also improvise our time complexity by implementing this on hardware accelerated GPU. We also plan to improve our Ontology by adding `experiences' to malware instances.

Moreover, we demonstrated that the gated-RRAM neuromorphic architecture serves as a versatile implentation of a hardware neural network.
The weights can be trained off-chip using software models and programmed during run-time.
Additionally, more biologically inspired algorithms can be applied to train the synaptic devices on-chip.
The gated-RRAM architecture is able to tune is learned model through different training algorithms and through different device characteristic such as conductance distribution and noise from device write variance.
However, we observed that this versatile model is still susceptible to an unknown adversarial attack. We have also demonstrated the ability to load (write) pre-trained weights to the RRAM neuromorphic architecture model.
Therefore, a software network trained on adversarial models can be implemented using the RRAM neurormorphic architectures allowing it to be deployed in low-power real-time applications while mitigating attacks from adversaries.
Additionally, we would like to explore methods for the chip to detect unknown adversarial attacks so that it may later train on them and identify adversarial inputs or sources.
Our goal is to develop a reconfigurable architecture that is robust to adversarial attacks and with the capabilities to be integrated in low-power IoT devices or applications.

%% file: Future.tex
\section{Future Work}

We aim to expand this work in the future and expand the scope of these techniques. Firstly, we will augment the malware data and then try to explore the performance of deep learning based techniques for malware detection. Secondly, we also look forward to generate stealthy malware data by adding complexity in the malware generated. The idea will be to evaluate the performance of deep learning and pattern recognition techniques over the stealthy malware data. We also plan to work on expanding the ontology of the malware classes and sub-classes based on the features extracted, to interact with the computers to make an efficient decisions. Moreover, our research will further investigate transfer learning of adversarial attacks for making an efficient RRAM based adversarial attack mitigation. Thus, our intent will be to make the RRAM-based neuromorphic architecture more robust against adversarial attacks.

%% file: rqe_main.bbl
\begin{thebibliography}{100}
\providecommand{\url}[1]{#1}
\csname url@samestyle\endcsname
\providecommand{\newblock}{\relax}
\providecommand{\bibinfo}[2]{#2}
\providecommand{\BIBentrySTDinterwordspacing}{\spaceskip=0pt\relax}
\providecommand{\BIBentryALTinterwordstretchfactor}{4}
\providecommand{\BIBentryALTinterwordspacing}{\spaceskip=\fontdimen2\font plus
\BIBentryALTinterwordstretchfactor\fontdimen3\font minus
  \fontdimen4\font\relax}
\providecommand{\BIBforeignlanguage}[2]{{%
\expandafter\ifx\csname l@#1\endcsname\relax
\typeout{** WARNING: IEEEtran.bst: No hyphenation pattern has been}%
\typeout{** loaded for the language `#1'. Using the pattern for}%
\typeout{** the default language instead.}%
\else
\language=\csname l@#1\endcsname
\fi
#2}}
\providecommand{\BIBdecl}{\relax}
\BIBdecl

\bibitem{Dhavlle_DATE'21}
A.~Dhavlle, S.~Shukla, S.~Rafatirad, H.~Homayoun, and S.~M.~P. Dinakarrao,
  ``Hmd-hardener: Adversarially robust and efficient hardware-assisted runtime
  malware detection,'' in \emph{Design, Automation and Test in Europe (DATE)},
  2021.

\bibitem{Dhavlle_ISCAS'21}
A.~Dhavlle, R.~Hassan, M.~Mittapalli, and S.~M.~P. D, ``Design of hardware
  trojans and its impact on cps systems: A comprehensive survey,'' in
  \emph{International Symposium on Circuits and Systems (ISCAS)}, 2021.

\bibitem{Meraj_ISCAS'21}
M.~M. Ahmed, A.~Dhavlle, N.~Mansoor, S.~M.~P. D, K.~Basu, and A.~Ganguly,
  ``What can a remote access hardware trojan do to a network-on-chip?'' in
  \emph{International Symposium on Circuits and Systems (ISCAS)}, 2021.

\bibitem{SMPD_DAC'19}
S.~M.~P. Dinakarrao, S.~Amberkar, S.~Bhat, A.~Dhavlle, H.~Sayadi, A.~Sasan,
  H.~Homayoun, and S.~Rafatirad, ``Adversarial attack on microarchitectural
  events based malware detectors,'' in \emph{Design Automation Conference
  (DAC)}, 2019.

\bibitem{Sanket_CASES'19}
S.~Shukla, G.~Kolhe, S.~M.~P. D, and S.~Rafatirad, ``Microarchitectural events
  and image processing-based hybrid approach for robust malware detection:
  Work-in-progress,'' in \emph{Proceedings of the International Conference on
  Compliers, Architectures and Synthesis for Embedded Systems Companion}, 2019.

\bibitem{Sanket_ICMLA'19}
S.~{Shukla}, G.~{Kolhe}, S.~M. {PD}, and S.~{Rafatirad}, ``Rnn-based classifier
  to detect stealthy malware using localized features and complex symbolic
  sequence,'' in \emph{2019 18th IEEE International Conference On Machine
  Learning And Applications (ICMLA)}, 2019, pp. 406--409.

\bibitem{Sanket_ICTAI'19}
S.~{Shukla}, G.~{Kolhe}, S.~M. {P D}, and S.~{Rafatirad}, ``Stealthy malware
  detection using rnn-based automated localized feature extraction and
  classifier,'' in \emph{2019 IEEE 31st International Conference on Tools with
  Artificial Intelligence (ICTAI)}, 2019, pp. 590--597.

\bibitem{Yarom_usenix_'14}
Y.~Yarom and K.~Falkner, ``Flush+reload: A high resolution, low noise, l3 cache
  side-channel attack,'' in \emph{USENIX Conference on Security}, 2014.

\bibitem{Gruss_dimva_'16}
D.~Gruss, C.~Maurice, K.~Wagner, and S.~Mangard, ``Flush+flush: A fast and
  stealthy cache attack,'' in \emph{Int. Conf. on Detection of Intrusions and
  Malware, and Vulnerability Assessment}, 2016.

\bibitem{Abhijitt_isqed'20}
A.~Dhavlle, R.~Mehta, S.~Rafatirad, H.~Homayoun, and S.~M.~P. Dinakarrao,
  ``Entropy-shield: Side-channel entropy maximization for timing-based
  side-channel attacks,'' in \emph{International Symposium on Quality
  Electronic Design (ISQED)}, 2020.

\bibitem{Dhavlle_TCAD'21}
A.~{Dhavlle}, S.~{Rafatirad}, K.~{Khasawneh}, H.~{Homayoun}, and S.~M.~P.
  {Dinakarrao}, ``Imitating functional operations for mitigating side-channel
  leakage,'' \emph{IEEE Transactions on Computer-Aided Design of Integrated
  Circuits and Systems}, pp. 1--1, 2021.

\bibitem{Meraj_AsianHOST'20}
M.~{Meraj Ahmed}, A.~{Dhavlle}, N.~{Mansoor}, P.~{Sutradhar}, S.~M. {Pudukotai
  Dinakarrao}, K.~{Basu}, and A.~{Ganguly}, ``Defense against on-chip trojans
  enabling traffic analysis attacks,'' in \emph{2020 Asian Hardware Oriented
  Security and Trust Symposium (AsianHOST)}, 2020, pp. 1--6.

\bibitem{Kolhe_GLSVLSI'19}
G.~Kolhe, S.~M. PD, S.~Rafatirad, H.~Mahmoodi, A.~Sasan, and H.~Homayoun, ``On
  custom lut-based obfuscation,'' in \emph{Proceedings of the 2019 on Great
  Lakes Symposium on VLSI}, 2019.

\bibitem{Kolhe_ICCAD'19}
G.~{Kolhe}, H.~M. {Kamali}, M.~{Naicker}, T.~D. {Sheaves}, H.~{Mahmoodi}, P.~D.
  {Sai Manoj}, H.~{Homayoun}, S.~{Rafatirad}, and A.~{Sasan}, ``Security and
  complexity analysis of lut-based obfuscation: From blueprint to reality,'' in
  \emph{2019 IEEE/ACM International Conference on Computer-Aided Design
  (ICCAD)}, 2019.

\bibitem{Hassan_ISQED'20}
R.~{Hassan}, G.~{Kolhe}, S.~{Rafatirad}, H.~{Homayoun}, and S.~M. {Dinakarrao},
  ``Satconda: Sat to sat-hard clause translator,'' in \emph{International
  Symposium on Quality Electronic Design (ISQED)}, 2020.

\bibitem{dac2021}
G.~Kolhe, S.~Salehi, T.~D. Sheaves, H.~Homayoun, S.~Rafatirad, M.~P.~D. Sai,
  and A.~Sasan, ``Securing hardware via dynamic obfuscation utilizing
  reconfigurable interconnect and logic blocks,'' in \emph{2021 58th ACM/IEEE
  Design Automation Conference (DAC)}, 2021, pp. 229--234.

\bibitem{dac2022_2}
G.~Kolhe, T.~D. Sheaves, K.~I. Gubbi, T.~Kadale, S.~Rafatirad, S.~M.~P.
  Dinakarrao, A.~Sasan, H.~Mahmoodi, and H.~Homayoun, ``Silicon validation of
  {LUT}-based logic-locked {IP} cores,'' in \emph{2022 59th ACM/IEEE Design
  Automation Conference (DAC)}, 2022.

\bibitem{dac2022}
G.~Kolhe, S.~Salehi, T.~D. Sheaves, H.~Homayoun, S.~Rafatirad, S.~M.~P.
  Dinakarrao, and A.~Sasan, ``{LOCK \& ROLL: Deep-Learning Power Side-Channel
  Attack Mitigation using Emerging Reconfigurable Devices and Logic Locking},''
  in \emph{2022 59th ACM/IEEE Design Automation Conference (DAC)}, 2022.

\bibitem{myjournal}
\BIBentryALTinterwordspacing
G.~Kolhe, T.~D. Sheaves, S.~M.~P. D, H.~Mahmoodi, S.~Rafatirad, A.~Sasan, and
  H.~Homayoun, ``Breaking the design and security trade-off of look-up
  table-based obfuscation,'' \emph{ACM Trans. Des. Autom. Electron. Syst.}, jan
  2022. [Online]. Available: \url{https://doi.org/10.1145/3510421}
\BIBentrySTDinterwordspacing

\bibitem{frontiers}
\BIBentryALTinterwordspacing
Z.~Chen, L.~Zhang, G.~Kolhe, H.~M. Kamali, S.~Rafatirad, S.~M.
  Pudukotai~Dinakarrao, H.~Homayoun, C.-T. Lu, and L.~Zhao, ``Deep graph
  learning for circuit deobfuscation,'' May 2021. [Online]. Available:
  \url{https://www.ncbi.nlm.nih.gov/pmc/articles/PMC8184091/}
\BIBentrySTDinterwordspacing

\bibitem{DATE}
Z.~Chen, G.~Kolhe, S.~Rafatirad, C.-T. Lu, S.~Manoj~P.D., H.~Homayoun, and
  L.~Zhao, ``Estimating the circuit de-obfuscation runtime based on graph deep
  learning,'' in \emph{2020 Design, Automation Test in Europe Conference
  Exhibition (DATE)}, 2020, pp. 358--363.

\bibitem{raki1}
R.~Hassan, G.~Kolhe, S.~Rafatirad, H.~Homayoun, and S.~M.~P. Dinakarrao, ``A
  neural network-based cognitive obfuscation towards enhanced logic locking,''
  \emph{IEEE Transactions on Computer-Aided Design of Integrated Circuits and
  Systems}, pp. 1--1, 2021.

\bibitem{raki2}
R.~Hassan, G.~Kolhe, S.~Rafatirad, H.~Homayoun, and S.~M. Dinakarrao,
  ``{SATConda: SAT to SAT-Hard Clause Translator},'' in \emph{2020 21st
  International Symposium on Quality Electronic Design (ISQED)}, 2020, pp.
  155--160.

\bibitem{vivek1}
V.~V. Menon, G.~Kolhe, A.~Schmidt, J.~Monson, M.~French, Y.~Hu, P.~A. Beerel,
  and P.~Nuzzo, ``System-level framework for logic obfuscation with quantified
  metrics for evaluation,'' in \emph{2019 IEEE Cybersecurity Development
  (SecDev)}, 2019, pp. 89--100.

\bibitem{vivek2019}
V.~Menon, G.~Kolhe, A.~Schmidt, J.~Monson, M.~French, Y.~Hu, P.~A. Beerel, and
  P.~Nuzzo, ``Quantifying security and overheads for obfuscation of integrated
  circuits,'' in \emph{GOMACTech Conference}, 2019.

\bibitem{vivek2020}
V.~Menon, G.~Kolhe, J.~Fifty, A.~Schmidt, J.~Monson, M.~French, Y.~Hu, P.~A.
  Beerel, and P.~Nuzzo, ``Logic obfuscation: Modeling attack resiliency,'' in
  \emph{GOMACTech Conference}, 2020.

\bibitem{sign_based_technique_1}
G.~Jacob, H.~Debar, and E.~Filiol, ``Behavioral detection of malware: From a
  survey towards an established taxonomy,'' \emph{Journal in Computer
  Virology}, vol.~4, pp. 251--266, 08 2008.

\bibitem{sign_based_technique_2}
N.~{Patel}, A.~{Sasan}, and H.~{Homayoun}, ``Analyzing hardware based malware
  detectors,'' in \emph{2017 54th ACM/EDAC/IEEE Design Automation Conference
  (DAC)}, 2017, pp. 1--6.

\bibitem{sign_based_technique_3}
Q.~{Chen} and R.~A. {Bridges}, ``Automated behavioral analysis of malware: A
  case study of wannacry ransomware,'' in \emph{2017 16th IEEE International
  Conference on Machine Learning and Applications (ICMLA)}, 2017, pp. 454--460.

\bibitem{sign_based_technique_4}
A.~{Nazari}, N.~{Sehatbakhsh}, M.~{Alam}, A.~{Zajic}, and M.~{Prvulovic},
  ``Eddie: Em-based detection of deviations in program execution,'' in
  \emph{2017 ACM/IEEE 44th Annual International Symposium on Computer
  Architecture (ISCA)}, 2017, pp. 333--346.

\bibitem{self_driving_1}
J.~Stilgoe, ``Machine learning, social learning and the governance of
  self-driving cars,'' \emph{Social Studies of Science}, 11 2017.

\bibitem{self_driving_survey}
S.~Grigorescu, B.~Trasnea, T.~Cocias, and G.~Macesanu, ``A survey of deep
  learning techniques for autonomous driving,'' \emph{Journal of Field
  Robotics}, 11 2019.

\bibitem{shukla_wip}
S.~{Shukla} and et~al., ``Work-in-progress: Microarchitectural events and image
  processing-based hybrid approach for robust malware detection,'' in
  \emph{CASES}, 2019.

\bibitem{sanket_dac21}
S.~Shukla, P.~D. Sai~Manoj, G.~Kolhe, and S.~Rafatirad, ``On-device malware
  detection using performance-aware and robust collaborative learning,'' in
  \emph{2021 58th ACM/IEEE Design Automation Conference (DAC)}, 2021.

\bibitem{sreenitha_1}
S.~Kasarapu, S.~Shukla, R.~Hassan, A.~Sasan, H.~Homayoun, and S.~M. PD,
  ``Cad-fsl: Code-aware data generation based few-shot learning for efficient
  malware detection,'' in \emph{Proceedings of the Great Lakes Symposium on
  VLSI 2022}, ser. GLSVLSI '22.\hskip 1em plus 0.5em minus 0.4em\relax New
  York, NY, USA: Association for Computing Machinery, 2022.

\bibitem{szegedy2013intriguing}
C.~Szegedy, W.~Zaremba, I.~Sutskever, J.~Bruna, D.~Erhan, I.~Goodfellow, and
  R.~Fergus, ``Intriguing properties of neural networks,'' \emph{arXiv preprint
  arXiv:1312.6199}, 2013.

\bibitem{goodfellow2014explaining}
I.~J. Goodfellow, J.~Shlens, and C.~Szegedy, ``Explaining and harnessing
  adversarial examples,'' \emph{arXiv preprint arXiv:1412.6572}, 2014.

\bibitem{papernot2016limitations}
N.~Papernot, P.~McDaniel, S.~Jha, M.~Fredrikson, Z.~B. Celik, and A.~Swami,
  ``The limitations of deep learning in adversarial settings,'' in \emph{2016
  IEEE European symposium on security and privacy (EuroS\&P)}.\hskip 1em plus
  0.5em minus 0.4em\relax IEEE, 2016, pp. 372--387.

\bibitem{mnist}
\BIBentryALTinterwordspacing
Y.~LeCun and C.~Cortes, ``{MNIST} handwritten digit database,'' 2010. [Online].
  Available: \url{http://yann.lecun.com/exdb/mnist/}
\BIBentrySTDinterwordspacing

\bibitem{5_Yoo_2004}
I.~Yoo, ``Visualizing windows executable viruses using self-organizing maps,''
  in \emph{Proceedings of the 2004 ACM Workshop on Visualization and Data
  Mining for Computer Security}, ser. VizSEC/DMSEC '04, 2004.

\bibitem{6_quist}
D.~A. {Quist} and L.~M. {Liebrock}, ``Visualizing compiled executables for
  malware analysis,'' in \emph{2009 6th International Workshop on Visualization
  for Cyber Security}, 2009.

\bibitem{7_trinius}
P.~{Trinius}, T.~{Holz}, J.~{Göbel}, and F.~C. {Freiling}, ``Visual analysis
  of malware behavior using treemaps and thread graphs,'' in \emph{2009 6th
  International Workshop on Visualization for Cyber Security}, 2009.

\bibitem{8_Goodall}
J.~R. Goodall, H.~Radwan, and L.~Halseth, ``Visual analysis of code security,''
  in \emph{Proceedings of the Seventh International Symposium on Visualization
  for Cyber Security}, ser. VizSec '10, 2010.

\bibitem{9_Conti}
G.~Conti, S.~Bratus, A.~Shubina, A.~Lichtenberg, B.~Sangster, and M.~Supan, ``A
  visual study of primitive binary fragment types,'' 2010.

\bibitem{10_Conti}
G.~Conti, S.~Bratus, A.~Shubina, B.~Sangster, R.~Ragsdale, M.~Supan,
  A.~Lichtenberg, and R.~Perez-Alemany, ``Automated mapping of large binary
  objects using primitive fragment type classification,'' \emph{Digit.
  Investig.}, vol.~7, Aug. 2010.

\bibitem{11_Karim05malwarephylogeny}
M.~E. Karim, A.~Walenstein, A.~Lakhotia, and L.~Parida, ``Malware phylogeny
  generation using permutations of code,'' \emph{JOURNAL IN COMPUTER VIROLOGY},
  vol.~1, pp. 13--23, 2005.

\bibitem{12_Kolter}
J.~Z. Kolter and M.~A. Maloof, ``Learning to detect and classify malicious
  executables in the wild,'' \emph{J. Mach. Learn. Res.}, vol.~7, Dec. 2006.

\bibitem{13_Gao}
D.~Gao, M.~K. Reiter, and D.~Song, ``Binhunt: Automatically finding semantic
  differences in binary programs,'' in \emph{Proceedings of the 10th
  International Conference on Information and Communications Security}, ser.
  ICICS '08, 2008.

\bibitem{14_Tian}
R.~{Tian}, L.~M. {Batten}, and S.~C. {Versteeg}, ``Function length as a tool
  for malware classification,'' in \emph{2008 3rd International Conference on
  Malicious and Unwanted Software (MALWARE)}, Oct 2008.

\bibitem{15_tian}
R.~{Tian}, L.~{Batten}, R.~{Islam}, and S.~{Versteeg}, ``An automated
  classification system based on the strings of trojan and virus families,'' in
  \emph{2009 4th International Conference on Malicious and Unwanted Software
  (MALWARE)}, 2009.

\bibitem{16_Tian}
R.~{Islam}, R.~{Tian}, L.~{Batten}, and S.~{Versteeg}, ``Classification of
  malware based on string and function feature selection,'' in \emph{2010
  Second Cybercrime and Trustworthy Computing Workshop}, 2010.

\bibitem{17_Gheorghescu2006ANAV}
M.~Gheorghescu, ``An automated virus classification system,'' 2006.

\bibitem{18_park}
Y.~Park, D.~Reeves, V.~Mulukutla, and B.~Sundaravel, ``Fast malware
  classification by automated behavioral graph matching,'' in \emph{Proceedings
  of the Sixth Annual Workshop on Cyber Security and Information Intelligence
  Research}, ser. CSIIRW '10, 2010.

\bibitem{19_Bailey:2007}
M.~Bailey, J.~Oberheide, J.~Andersen, Z.~M. Mao, F.~Jahanian, and J.~Nazario,
  ``Automated classification and analysis of internet malware,'' in
  \emph{Proceedings of the 10th International Conference on Recent Advances in
  Intrusion Detection}, ser. RAID'07, 2007.

\bibitem{20_Bayer_scalable}
U.~Bayer, P.~M. Comparetti, C.~Hlauschek, C.~Kruegel, and E.~Kirda, ``Scalable,
  behavior-based malware clustering.''

\bibitem{21_Rieck:2008}
K.~Rieck, T.~Holz, C.~Willems, P.~D\"{u}ssel, and P.~Laskov, ``Learning and
  classification of malware behavior,'' in \emph{Proceedings of the 5th
  International Conference on Detection of Intrusions and Malware, and
  Vulnerability Assessment}, ser. DIMVA '08, 2008.

\bibitem{Abhijitt_1}
S.~M.~P. Dinakarrao, S.~Amberkar, S.~Bhat, A.~Dhavlle, H.~Sayadi, A.~Sasan,
  H.~Homayoun, and S.~Rafatirad, ``Adversarial attack on microarchitectural
  events based malware detectors,'' in \emph{Proceedings of the 56th Annual
  Design Automation Conference 2019}, 2019.

\bibitem{Abhijitt_2}
F.~Brasser, L.~Davi, A.~Dhavlle, T.~Frassetto, S.~M.~P. Dinakarrao,
  S.~Rafatirad, A.-R. Sadeghi, A.~Sasan, H.~Sayadi, S.~Zeitouni, and
  H.~Homayoun, ``Advances and throwbacks in hardware-assisted security: Special
  session,'' in \emph{Proceedings of the International Conference on Compilers,
  Architecture and Synthesis for Embedded Systems}, 2018.

\bibitem{Abhijitt_3}
A.~Dhavlle, R.~Mehta, S.~Rafatirad, H.~Homayoun, and S.~M.
  Pudukotai~Dinakarrao, ``Entropy-shield:side-channel entropy maximization for
  timing-based side-channel attacks,'' in \emph{2020 21st International
  Symposium on Quality Electronic Design (ISQED)}, 2020, pp. 161--166.

\bibitem{Abhijitt_4}
A.~Dhavlle, R.~Hassan, M.~Mittapalli, and S.~M.~P. Dinakarrao, ``Design of
  hardware trojans and its impact on cps systems: A comprehensive survey,'' in
  \emph{International Symposium on Circuits and Systems (ISCAS)}, 2021, pp.
  1--5.

\bibitem{Abhijitt_5}
A.~Dhavlle, S.~Rafatirad, K.~Khasawneh, H.~Homayoun, and S.~M.~P. Dinakarrao,
  ``Imitating functional operations for mitigating side-channel leakage,''
  \emph{IEEE Transactions on Computer-Aided Design of Integrated Circuits and
  Systems}, vol.~41, no.~4, pp. 868--881, 2022.

\bibitem{Abhijitt_6}
A.~Dhavlle, S.~Shukla, S.~Rafatirad, H.~Homayoun, and S.~M.
  Pudukotai~Dinakarrao, ``Hmd-hardener: Adversarially robust and efficient
  hardware-assisted runtime malware detection,'' in \emph{Design, Automation \&
  Test in Europe Conference \& Exhibition (DATE)}, 2021, pp. 1769--1774.

\bibitem{Abhijitt_7}
M.~Meraj~Ahmed, A.~Dhavlle, N.~Mansoor, P.~Sutradhar, S.~M.
  Pudukotai~Dinakarrao, K.~Basu, and A.~Ganguly, ``Defense against on-chip
  trojans enabling traffic analysis attacks,'' in \emph{Asian Hardware Oriented
  Security and Trust Symposium (AsianHOST)}, 2020, pp. 1--6.

\bibitem{Hall_weka'09}
\BIBentryALTinterwordspacing
M.~Hall, E.~Frank, G.~Holmes, B.~Pfahringer, P.~Reutemann, and I.~H. Witten,
  ``The weka data mining software: An update,'' \emph{SIGKDD Explor. Newsl.},
  vol.~11, no.~1, pp. 10--18, Nov. 2009. [Online]. Available:
  \url{http://doi.acm.org/10.1145/1656274.1656278}
\BIBentrySTDinterwordspacing

\bibitem{naive_bayes}
\emph{https://towardsdatascience.com/introduction-to-naive-bayes-classification-4cffabb1ae54},
  last accessed: 14th May 2019.

\bibitem{random_forest}
\emph{https://towardsdatascience.com/the-random-forest-algorithm-d457d499ffcd},
  last accessed: 14th May 2019.

\bibitem{logistic_regression}
\emph{https://ml-cheatsheet.readthedocs.io/en/latest/logistic\_regression.html},
  last accessed: 14th May 2019.

\bibitem{protege}
M.~A. Musen, ``The prot\'{E}g\'{E} project: A look back and a look forward,''
  \emph{AI Matters}.

\bibitem{sayadi-CF18}
H.~Sayadi, S.~M.~P. D, A.~Houmansadr, S.~Rafatirad, and H.~Homayoun,
  ``Comprehensive assessment of run-time hardware-supported malware detection
  using general and ensemble learning,'' in \emph{Proceedings of the 15th ACM
  International Conference on Computing Frontiers}, 2018.

\bibitem{DAC4'17}
N.~Patel, A.~Sasan, and H.~Homayoun, ``Analyzing hardware based malware
  detectors,'' in \emph{ACM/EDAA/IEEE Design Automation Conference}, 2017.

\bibitem{Demme'13}
J.~Demme, M.~Maycock, J.~Schmitz, A.~Tang, A.~Waksman, S.~Sethumadhavan, and
  S.~Stolfo, ``On the feasibility of online malware detection with performance
  counters,'' \emph{SIGARCH Comput. Archit. News}, vol.~41, no.~3, pp.
  559--570, Jun 2013.

\bibitem{Sanket_DAC'21}
S.~Shukla, G.~Kolhe, S.~M.~P. D, and S.~Rafatirad, ``On-device malware
  detection using performance-aware and robust collaborative learning,'' in
  \emph{Design Automation Conference (DAC)}, 2021.

\bibitem{Rootkit-Singh}
S.~Baljit and et~al., ``On the detection of kernel-level rootkits using
  hardware performance counters,'' in \emph{ACM AsiaCCS'17}, 2017.

\bibitem{Bahador'14}
M.~B. Bahador, M.~Abadi, and A.~Tajoddin, ``{HPCMalHunter}: Behavioral malware
  detection using hardware performance counters and singular value
  decomposition,'' in \emph{Int. Conf. on Computer and Knowledge Engineering},
  2014.

\bibitem{Jacob'08}
G.~Jacob, H.~Debar, and E.~Filiol, ``Behavioral detection of malware: from a
  survey towards an established taxonomy,'' \emph{Journal in Computer
  Virology}, vol.~4, no.~3, pp. 251--266, Aug 2008.

\bibitem{Nelson2008ExploitingML}
B.~Nelson, M.~Barreno, F.~J. Chi, A.~Joseph, B.~I.~P. Rubinstein, U.~Saini,
  C.~Sutton, J.~Tygar, and K.~Xia, ``Exploiting machine learning to subvert
  your spam filter,'' in \emph{LEET}, 2008.

\bibitem{rubinstein2009antidote}
B.~I. Rubinstein, B.~Nelson, L.~Huang, A.~D. Joseph, S.-h. Lau, S.~Rao,
  N.~Taft, and J.~D. Tygar, ``Antidote: understanding and defending against
  poisoning of anomaly detectors,'' in \emph{Proceedings of the 9th ACM SIGCOMM
  Conference on Internet Measurement}, 2009, pp. 1--14.

\bibitem{biggio2012poisoning}
B.~Biggio, B.~Nelson, and P.~Laskov, ``Poisoning attacks against support vector
  machines,'' \emph{arXiv preprint arXiv:1206.6389}, 2012.

\bibitem{munoz2017towards}
L.~Mu{\~n}oz-Gonz{\'a}lez, B.~Biggio, A.~Demontis, A.~Paudice, V.~Wongrassamee,
  E.~C. Lupu, and F.~Roli, ``Towards poisoning of deep learning algorithms with
  back-gradient optimization,'' in \emph{Proceedings of the 10th ACM Workshop
  on Artificial Intelligence and Security}, 2017, pp. 27--38.

\bibitem{sshukla4_rram}
S.~Barve, S.~Shukla, S.~M.~P. Dinakarrao, and R.~Jha, \emph{Adversarial Attack
  Mitigation Approaches Using RRAM-Neuromorphic Architectures}.\hskip 1em plus
  0.5em minus 0.4em\relax New York, NY, USA: Association for Computing
  Machinery, 2021.

\bibitem{Papernot2016DistillationAA}
N.~Papernot, P.~McDaniel, X.~Wu, S.~Jha, and A.~Swami, ``Distillation as a
  defense to adversarial perturbations against deep neural networks,''
  \emph{2016 IEEE Symposium on Security and Privacy (SP)}, pp. 582--597, 2016.

\bibitem{meng2017magnet}
D.~Meng and H.~Chen, ``Magnet: a two-pronged defense against adversarial
  examples,'' in \emph{Proceedings of the 2017 ACM SIGSAC conference on
  computer and communications security}, 2017, pp. 135--147.

\bibitem{Abhijitt_8}
A.~Dhavlle and S.~M. Pudukotai~Dinakarrao, ``A comprehensive review of ml-based
  time-series and signal processing techniques and their hardware
  implementations,'' in \emph{International Green and Sustainable Computing
  Workshops (IGSC)}, 2020, pp. 1--8.

\bibitem{Abhijitt_9}
S.~Bavikadi, A.~Dhavlle, A.~Ganguly, A.~Haridass, H.~Hendy, C.~Merkel, V.~J.
  Reddi, P.~R. Sutradhar, A.~Joseph, and S.~M. Pudukotai~Dinakarrao, ``A survey
  on machine learning accelerators and evolutionary hardware platforms,''
  \emph{IEEE Design \& Test}, vol.~39, no.~3, pp. 91--116, 2022.

\bibitem{Abhijitt_10}
A.~Dhavlle, S.~Rafatirad, H.~Homayoun, and S.~M.~P. Dinakarrao, ``Cr-spectre:
  Defense-aware rop injected code-reuse based dynamic spectre,'' in
  \emph{Design, Automation \& Test in Europe Conference \& Exhibition (DATE)},
  2022, pp. 508--513.

\bibitem{Abhijitt_12}
M.~M. Ahmed, A.~Dhavlle, N.~Mansoor, S.~M.~P. Dinakarrao, K.~Basu, and
  A.~Ganguly, ``What can a remote access hardware trojan do to a
  network-on-chip?'' in \emph{International Symposium on Circuits and Systems
  (ISCAS)}, 2021, pp. 1--5.

\bibitem{Abhijitt_13}
A.~Dhavlle and S.~Shukla, ``A novel malware detection mechanism based on
  features extracted from converted malware binary images,'' \emph{CoRR}, vol.
  abs/2104.06652, 2021.

\bibitem{Abhijitt_14}
``Table of contents,'' \emph{IEEE Transactions on Computer-Aided Design of
  Integrated Circuits and Systems}, vol.~41, no.~4, pp. C1--C4, 2022.

\bibitem{bhattacharjee2020rethinking}
A.~Bhattacharjee and P.~Panda, ``Rethinking non-idealities in memristive
  crossbars for adversarial robustness in neural networks,'' \emph{arXiv
  preprint arXiv:2008.11298}, 2020.

\bibitem{kim2019neuromorphic}
E.~Kim, J.~Yarnall, P.~Shah, and G.~T. Kenyon, ``A neuromorphic sparse coding
  defense to adversarial images,'' in \emph{Proceedings of the International
  Conference on Neuromorphic Systems}, 2019, pp. 1--8.

\bibitem{Kurakin2017AdversarialEI}
A.~Kurakin, I.~Goodfellow, and S.~Bengio, ``Adversarial examples in the
  physical world,'' \emph{ArXiv}, vol. abs/1607.02533, 2017.

\bibitem{Dong2018BoostingAA}
Y.~Dong, F.~Liao, T.~Pang, H.~Su, J.~Zhu, X.~Hu, and J.~Li, ``Boosting
  adversarial attacks with momentum,'' \emph{2018 IEEE/CVF Conference on
  Computer Vision and Pattern Recognition}, pp. 9185--9193, 2018.

\bibitem{MoosaviDezfooli2016DeepFoolAS}
S.-M. Moosavi-Dezfooli, A.~Fawzi, and P.~Frossard, ``Deepfool: A simple and
  accurate method to fool deep neural networks,'' \emph{2016 IEEE Conference on
  Computer Vision and Pattern Recognition (CVPR)}, pp. 2574--2582, 2016.

\bibitem{shaham2015understanding}
U.~Shaham, Y.~Yamada, and S.~Negahban, ``Understanding adversarial training:
  Increasing local stability of neural nets through robust optimization,''
  \emph{arXiv preprint arXiv:1511.05432}, 2015.

\bibitem{Carlini2017TowardsET}
N.~Carlini and D.~A. Wagner, ``Towards evaluating the robustness of neural
  networks,'' \emph{2017 IEEE Symposium on Security and Privacy (SP)}, pp.
  39--57, 2017.

\bibitem{hinton2015distilling}
G.~Hinton, O.~Vinyals, and J.~Dean, ``Distilling the knowledge in a neural
  network,'' \emph{arXiv preprint arXiv:1503.02531}, 2015.

\bibitem{Grosse2017OnT}
K.~Grosse, P.~Manoharan, N.~Papernot, M.~Backes, and P.~McDaniel, ``On the
  (statistical) detection of adversarial examples,'' \emph{ArXiv}, vol.
  abs/1702.06280, 2017.

\bibitem{Metzen2017OnDA}
J.~H. Metzen, T.~Genewein, V.~Fischer, and B.~Bischoff, ``On detecting
  adversarial perturbations,'' \emph{ArXiv}, vol. abs/1702.04267, 2017.

\bibitem{STLT}
T.~Bailey, A.~Ford, S.~Barve, J.~Wells, and R.~Jha, ``Development of a
  short-term to long-term supervised spiking neural network processor,''
  \emph{IEEE Transactions on Very Large Scale Integration (VLSI) Systems}, vol.
  doi: 10.1109/TVLSI.2020.3013810.

\bibitem{GSD}
A.~Jones and R.~Jha, ``A compact gated-synapse model for neuromorphic
  circuits,'' \emph{IEEE Transactions on Computer-Aided Design of Integrated
  Circuits and Systems}, vol. 10.1109/TCAD.2020.3028534.

\bibitem{3T-RRAM}
E.~Herrmann, A.~Rush, T.~Bailey, and R.~Jha, ``Gate controlled three-terminal
  metal oxide memristor,'' \emph{IEEE Electron Device Letters}, vol. vol. 39,
  no. 4, pp. 500-503, April 2018.

\bibitem{cleverhans}
N.~Papernot, F.~Faghri, N.~Carlini, I.~Goodfellow, R.~Feinman, A.~Kurakin,
  C.~Xie, Y.~Sharma, T.~Brown, A.~Roy, A.~Matyasko, V.~Behzadan,
  K.~Hambardzumyan, Z.~Zhang, Y.-L. Juang, Z.~Li, R.~Sheatsley, A.~Garg,
  J.~Uesato, W.~Gierke, Y.~Dong, D.~Berthelot, P.~Hendricks, J.~Rauber,
  R.~Long, and P.~McDaniel, ``Technical report on the cleverhans v2.1.0
  adversarial examples library,'' \emph{arXiv: Learning}, 2016.

\end{thebibliography}
